\documentclass[12pt]{article}
\usepackage{graphicx,cite,amsfonts,amssymb}%,draft}
\usepackage[body={6.75in,8.75in},nohead]{geometry} 
\def\baselinestretch{1.2}
 
   \newcommand\vpint{{\rm -}\kern -1.1em\int_{-\infty}^\infty}
   \newcommand{\reals}{{\rm I \mkern-2.5mu \nonscript\mkern-.5mu R}}
   \newcommand{\beq}{\begin{equation}}
   \newcommand{\eeq}{\end{equation}}
   \newcommand{\bea}{\begin{eqnarray}}
   \newcommand{\eea}{\end{eqnarray}}
   \newcommand{\beqstar}{\begin{displaymath}}
   \newcommand{\eeqstar}{\end{displaymath}}
   \newcommand{\beastar}{\begin{eqnarray*}}
   \newcommand{\eeastar}{\end{eqnarray*}}
   \newcommand{\bmath}[1]{\mbox{\boldmath{$#1$}}}
   \newcommand{\plus}{\mathop{\!+\!}}
   \newcommand{\minus}{\mathop{\!-\!}}
   \renewcommand{\t}{\theta}
   \newcommand{\te}{\vartheta_1}

   \newcommand{\tv}{\vartheta_4}
   
   \def\@@insvline#1#2{{\setbox0\hbox{\m@th$#1\rm I$} \rlap{\m@th$#1 \mkern5mu
                        \vrule height.92\ht0 depth-.05\ht0 width.09\ht0 $}
                        {\rm #2} }}
   \newcommand{\ie}{\mbox{\it i.e. \/}}

   \newcommand{\real}{\mathop{\it {\Re}e}}
   \newcommand{\imag}{\mathop{\it {\Im}m}}
   \newcommand{\p}{\mbox{p}}
   
   \newcommand{\boltzmannweight}[7]{W^{#6 #7} \! \! \left( \left. \!
                                    \begin{array}{cc} \! #1 \! & \! #2 \! \\
                                    \! #3 \! & \! #4 \! \end{array}  \right|
                                    #5 \right)}

   \newcommand{\journal}[6]{#1: #2, \textit{#3} \textbf{#4} (#5), #6.}
         \newcommand{\CMP}{Commun. Math. Phys.}
         
         \newcommand{\IJMP}{Int. J. Mod. Phys.}
         \newcommand{\JMP}{J. Math. Phys.}
         \newcommand{\JP}{J. Phys.}
         \newcommand{\JSP}{J. Stat. Phys.}
         \newcommand{\NP}{Nucl. Phys.}
         \newcommand{\PHYSLETT}{Phys. Lett.}

         \newcommand{\PHYSREVLETT}{Phys. Rev. Lett.}

   \newcommand{\book}[3]{#1: \textit{#2}, #3.}

\renewcommand{\thefootnote}{\fnsymbol{footnote}}

\renewcommand{\title}[1]{\bigskip\bigskip\Large\bf #1\bigskip\bigskip\\}
\renewcommand{\author}[1]{\large\rm #1\\ \bigskip}
\newcommand{\address}[1]{{\normalsize\it #1\\}\bigskip}

\begin{document}

%%%%%%%%%%%%%%%%%%%%%%%%%%%
% Title Page and Abstract %
%%%%%%%%%%%%%%%%%%%%%%%%%%%
\vskip 0.5cm
\begin{flushright}
{{hep-th/0205238}}
\end{flushright}
\vskip 2.4cm

\vglue .3 cm

\begin{center}
\title{Excited State TBA for the $\phi_{2,1}$ perturbed ${\cal M}_{3,5}$ model}

\author{R. M. Ellem,
        V. V. Bazhanov\footnote[3]{email:
                {\tt Vladimir.Bazhanov@anu.edu.au}}}

\address{Department of Theoretical Physics, RSPhysSE, IAS,\\
	Australian National University, Canberra, ACT 0200, Australia.}
\end{center}

\setcounter{footnote}{0}
\vspace{5mm}

\begin{abstract}
We examine some excited state energies in the non-unitary integrable quantum
field theory (IQFT) obtained from the perturbation of the minimal conformal 
field theory (CFT) model $\mathcal{M}_{3,5}$ by its operator $\phi_{2,1}$.
Using the correspondence of this IQFT  to the scaling limit of the 
dilute $A_{2}$ lattice model (in a particular regime) we derive the  
functional equations for the QFT commuting transfer matrices. 
These functional equations 
can be transformed to a closed set of TBA-like integral equations 
which determine the excited state energies in the finite-size system. In 
particular, we explicitly construct these equations for  
the ground state and two lowest excited states. Numerical results for
the associated energy gaps are compared 
with those obtained by the truncated conformal space approach (TCSA).
\end{abstract}

\newpage

%%%%%%%%%%%%%%%%%%%%%%
% Start of Main Text %
%%%%%%%%%%%%%%%%%%%%%%
\renewcommand{\thefootnote}{\arabic{footnote}}
\section{Introduction}
\setcounter{equation}{0}
The problem of calculating the excited state energies in finite volume 
integrable quantum field theories (IQFT's) has recently received 
much attention \cite{BazhanovLukyanovZamolodchikov4,DoreyTateo1,
FioravantiMariottiniQuattriniRavanini,DestrideVega,DoreyTateo2,Fendley}. 
In particular, in reference \cite{BazhanovLukyanovZamolodchikov4} it was 
shown that the functional equations for the IQFT ``commuting transfer-matrices'' 
introduced in \cite{BazhanovLukyanovZamolodchikov1} can be transformed 
\cite{KlumperPearce1,KlumperPearce2} into integral equations which generalise 
the standard ground state thermodynamic Bethe Ansatz (TBA) equations to 
excited states. The results of \cite{BazhanovLukyanovZamolodchikov4} apply to 
massive IQFT's obtained from $\phi_{1,3}$ perturbations of the minimal 
conformal field theory (CFT) models, which are related to the $U_{q}(
\widehat{sl}(2))$ quantum algebra. To include IQFT's related 
to $\phi_{1,2}$ or $\phi_{2,1}$ perturbations of these minimal CFT models,
one has to generalise this approach to the case of the $q$-deformed twisted
Kac-Moody algebra $A^{(2)}_2$. Some results in this direction were obtained
in \cite{FioravantiRavaniniStanishkov1,BHK}. 

In this paper we consider the IQFT obtained from the perturbation of the 
non-unitary ($c=-3/5$) minimal CFT  ${\cal M}_{3,5}$ by its operator  
$\phi_{2,1}$ with conformal dimensions $\Delta=\overline{\Delta}=3/4$.  
It is known that the particle spectrum of this model consists of three 
kinks of the same mass which interpolate between two degenerate vacuum 
states. The exact  $S$-matrix of these kinks was conjectured by Smirnov 
\cite{Smirnov1}. 
A numerical study of the energy spectrum of this model was performed in 
\cite{Mussardo2} using the Truncated Conformal Space Approach (TCSA)  
\cite{YurovZamolodchikov}. The TBA equations
for the ground state energy were originally conjectured in \cite{Ravanini3},
and more recently, were derived in \cite{EllemBazhanov1} from the exact 
$S$-matrix of the model. In this paper we derive the functional relation 
for the eigenvalues of the ``QFT transfer matrices'' which is, in principle, 
sufficient to determine the whole energy spectrum of the  model. In 
doing this we do not directly follow the route of works
\cite{BazhanovLukyanovZamolodchikov4,BazhanovLukyanovZamolodchikov1,
BazhanovLukyanovZamolodchikov2}, but instead obtain the required  
functional relation as the scaling limit of the associated lattice model
results. Fortunately, a suitable solvable lattice model whose scaling limit 
describes the above IQFT is known. It is a particular case of the 
off-critical dilute-$A_2$ lattice model  \cite{WarnaarNienhuisSeaton} in a 
certain regime. 
 
The organisation of the paper is as follows. In Section~2, we briefly review
the related results of \cite{EllemBazhanov1} where the TBA integral
equations for the ground state energy of the model were derived using the 
conventional TBA approach \cite{Yang2,Zamolodchikov5}. The massive field
theory limit of the related dilute-$A_2$ lattice model is considered in 
Section~3. In Section~4, we present the functional relation for the
QFT transfer matrices along with some numerical results on the positions of 
zeroes for the largest three eigenvalues of these transfer matrices.
Using these results we then derive the TBA-like integral equations which
correspond to the ground state and two lowest excited states energies of 
the finite-size QFT described above. Comparisons are made between the
two lowest excited state energy gaps computed from the above 
TBA-like integral equations and corresponding results \cite{Mussardo2} 
of the truncated conformal space approach (TCSA).

\section{The $S$-Matrix and Ground state TBA Equation}
\setcounter{equation}{0}

Consider the massive QFT described by the action
\beq
\mathcal{A}=\mathcal{A}_{\mathcal{M}_{3,5}}+g\int\phi_{2,1}\, d^2\! x 
\label{action}
\eeq
where ${\cal A}_{\mathcal{M}_{3,5}}$ represents the action of the $c=-3/5$ 
minimal non-unitary CFT ${\cal M}_{3,5}$, the operator $\phi_{2,1}$ has 
conformal dimensions $(3/4,3/4)$, and $g$ is a coupling constant of dimension 
$({\rm mass})^{1/2}$. It is known \cite{Zamolodchikov2}, that this action 
defines an integrable QFT in the sense that it possesses an infinite number 
of non-trivial integrals of motion. The particle spectrum of the model 
consists of three kinks (of the same mass $m$) interpolating between two 
degenerate vacuum states (labelled $0$ and $1$). The precise relationship
between the perturbation parameter $g$ and the kink mass $m$ is \cite{Fateev}
\beq
g=\frac{i\,\Gamma\left(\frac{5}{12}\right)\Gamma\left(\frac{1}{3}\right)}{
2\sqrt{3} \pi\Gamma\left(\frac{7}{12}\right)\Gamma\left(\frac{2}{3}\right)}
m^{\frac{1}{2}} \approx 0.25300091957\ldots i\, m^{\frac{1}{2}} .
\eeq
The exact kink-kink $S$-matrix (as first conjectured in \cite{Smirnov1}) can 
be written in the form \cite{Mussardo2}
\beq
S_{\alpha \beta}^{\gamma \delta} ( \theta ) = \left( \frac{\rho_{
\gamma} \rho_{\delta}}{\rho_{\alpha} \rho_{\beta}} \right)^{-
\frac{\theta}{2 \pi i}} R ( \theta )\,\, W \! \! \left( \left.
\begin{array}{cc} \alpha & \delta \\ \gamma & \beta \end{array}
\right| \varkappa \theta \right), \qquad \alpha,\beta,\gamma,\delta
\in \{ 0,1 \}  \label{eq:smatrix}
\eeq
where $\theta$ is the physical rapidity variable, the parameters $\varkappa= 
-\frac{3i}{5}$, $\;\rho_0=2\cos\mu\;$ and $\;\rho_1=1$, and the normalisation 
factor is
\beq
R( \theta ) = \frac{\sin \left( \frac{2 \pi}{5} - \frac{3 i \theta}{5} \right)
\sin \left( \frac{\pi}{5} - \frac{3 i \theta}{5} \right)}{\sin \left(
\frac{2 \pi}{5} + \frac{3 i \theta}{5} \right) \sin \left( \frac{\pi}{5} +
\frac{3 i \theta}{5} \right)} .
\label{eq:normal}
\eeq
Here, the functions $W(...)$ denote the Boltzmann weights of the critical 
hard hexagon lattice model \cite{Baxter4}  (in an unphysical regime). With a 
suitable normalisation these Boltzmann weights can be expressed as
\bea
\boltzmannweight{0}{0}{0}{0}{u}{}{} & = & \frac{\sin \mu \,\sin (2\mu
\plus u)}{\sin 2\mu\, \sin (\mu\minus u)}  \nonumber \\
\boltzmannweight{0}{0}{1}{0}{u}{}{} & = & \boltzmannweight{0}{1}{0}{0}{u}{}{} =
\left[\frac{\sin\mu}{\sin2\mu}\right]^{\frac{1}{2}}{\sin u\over\sin(\mu\minus
u)} \nonumber \\
\boltzmannweight{1}{0}{0}{0}{u}{}{} & = & \boltzmannweight{0}{0}{0}{1}{u}{}{} =
1  \label{BW} \\
\boltzmannweight{0}{1}{1}{0}{u}{}{} & = & {\sin \mu \over \sin 2\mu}
\frac{\sin (2\mu\minus u)}{\sin(\mu\minus u)} \nonumber \\
\boltzmannweight{1}{0}{0}{1}{u}{}{} & = & \frac{\sin (\mu\plus u)}{\sin
(\mu\minus u)} \nonumber
\eea
where $u\! =\!\varkappa\theta$ is the lattice spectral parameter and 
$\mu\! =\! 3\pi/5$ is the crossing parameter. 

The $S$-matrix (\ref{eq:smatrix}) defines a factorized scattering theory. This
implies that a system of $N$ kinks distributed along a large spatial circle of 
length $L$ can be analysed via the usual Bethe Ansatz (BA) approach. The rapidities 
$\theta_{1},\ldots ,\theta_{N}$ of the respective kinks are consequently 
constrained by the Bethe-Yang equations \cite{Zamolodchikov5}
%%% The $S$-matrix (\ref{eq:smatrix}) defines a factorized scattering
%%%theory which 
%%% can be analysed via the Bethe Ansatz. The rapidities $\theta_{1},
%%% \ldots ,\theta_{N}$ of a system of $N$ kinks distributed
%%% along a large spatial circle of length $L$ are constrained by
%%% the Bethe-Yang equations \cite{Zamolodchikov5}
\beq
e^{imL \sinh  \theta_{k}} \Lambda(\theta_{k};\theta_{1},\ldots ,
\theta_{N} ) = -1 , \; \; \; \; \; \; \; k=1, \ldots ,N \label{eq:betye}
\eeq
where $\Lambda(\theta;\theta_{1}, \ldots ,\theta_{N} )$ are the
eigenvalues of the ``scattering transfer matrix''
\beq
\bmath{T}(\theta;\theta_{1},\ldots ,\theta_{N} )_{\left\{ \alpha
\right\}}^{\left\{ \beta \right\}} = \prod_{i=1}^{N} S_{\beta_{i}
\alpha_{i+1}}^{\alpha_{i} \beta_{i+1}} ( \theta - \theta_{i} ) .\label{tmatrix}
\eeq
{}From equation (\ref{eq:smatrix}) it is obvious that this is just 
the transfer matrix of the inhomogeneous lattice model with the 
Boltzmann weights given by (\ref{BW}).
The eigenvalues of this lattice model were found in 
\cite{BazhanovReshetikhin1} using the analytic Bethe Ansatz approach.
This result (together with some assumptions on the string structure of 
the associated Bethe-Ansatz equations) enables one to follow the standard
TBA procedure \cite{EllemBazhanov1}. Below we briefly review these
calculations. 

The states of the system 
in the thermodynamic limit $N\rightarrow\infty$, $L\rightarrow\infty$,
 are specified  by the densities of the rapidity distributions for 
the kinks and quasi-particles arising in the diagonalization of the 
transfer matrix (\ref{tmatrix}). As shown in \cite{EllemBazhanov1},
only one type of quasi-particle is required in our case. The densities 
of the rapidity distributions for the kink and the quasi-particle (and 
the associated densities of ``holes'' in these distributions) are then 
determined by the following integral equations \cite{EllemBazhanov1}
\beq
{m\over2\pi} \delta_{j,0}\,\cosh\theta= \sigma_j(\theta)+\widetilde{
\sigma}_j(\theta)+ \sum_{k=0}^1 \int_{-\infty}^\infty \Phi_{j,k}(\theta
-\theta') \,\,\sigma_k(\theta') \,\, d\theta', \qquad j=0,1. 
\label{eq6:denseq1}
\eeq
where 
\beq
\Phi_{j,k}(\theta) =(\delta_{j,k}\minus\delta_{j,k-1}\minus\delta_{j,k+1})
\phi(\theta), \qquad\phi(\theta)=  
\frac{\sqrt{3}}{\pi} \frac{\sinh(2\theta)}{\sinh(3\theta)} .\label{phi}
\eeq      
Here $\sigma_0(\theta)$ denotes the density of the kink rapidity
distribution which is normalised as 
\begin{equation}
L \int_{- \infty}^{\infty} \sigma_0( \theta ) d \theta \! \!  =  \!
        \! N ,
\end{equation}
and $\widetilde{\sigma}_0(\theta)$ denotes the corresponding density of 
``holes'' in this distribution. Similarly, $\sigma_1(\theta)$  and 
$\widetilde{\sigma}_1(\theta)$ denote the rapidity and hole densities
for the quasi-particles. The density of states in the system  is  
determined by the usual combinatorial entropy \cite{Yang1} 
\beq
{\cal S}[\sigma_0(\theta),{\sigma}_1(\theta)] 
= L\sum_{k=0}^{1}\int_{-\infty}^{\infty}\left\{ \rule{0mm}{4mm}
[\sigma_{j}(\theta)\plus\tilde{\sigma}_{j}(\theta)]\ln[\sigma_{j}(\theta)\plus
\tilde{\sigma}_{j}(\theta)]\minus\sigma_{j}(\theta)\ln\sigma_{j}(\theta)\minus
\tilde{\sigma}_{j}(\theta)\ln\tilde{\sigma}_{j}(\theta)\right\}d\theta .\qquad
\label{eq6:entropy}
\eeq
The free energy of the system is then defined as the functional
\beq
{\cal F} [ \sigma_{0} ( \theta ), {\sigma}_{1} (\theta )] = {\cal E}
[ {\sigma}_0 ( \theta )] - \frac{1}{R} {\cal S} [ \sigma_{0} ( \theta ),
\sigma_1(\theta )] , \label{eq:fe}
\eeq
where $1/R$ acts as the temperature parameter, and the energy is 
\beq
{\cal E}[\sigma_0(\theta)]
 =\epsilon L+ m L \int_{-\infty}^{\infty} \cosh\theta \, \sigma_{0} ( \theta
) d \theta\  .\label{Eqft}
\eeq
Here, the parameter
\beq
\epsilon=m^2 \log(m R_0)/(8\pi)  \label{Epsqft}
\eeq
is the vacuum energy (per unit length) for the QFT
defined by the action (\ref{action}) \cite{Fateev,Ravanini3},
and $R_0$ is a non-universal ultraviolet cutoff parameter. 
Using the standard calculations, one obtains the equilibrium free energy
in the form 
\begin{equation}
f_{0}(mR)={{\cal F}\over L}= \epsilon -\frac{m}{2\pi R}\int_{-\infty}^{\infty}
\cosh\theta\,\log\left(\rule{0mm}{4mm}1+e^{-\varepsilon_0(\theta)}\,\right)
\,\, d\theta  \label{eq:minfe1}
\end{equation}
where the pseudo-energies
$$\varepsilon_j(\theta)=\log\left(\rule{0mm}{4mm}\widetilde{\sigma}_j(\theta)
/{\sigma}_j(\theta)\right)$$
are determined by the TBA integral equations \cite{Ravanini3,EllemBazhanov1}
\begin{equation}
\varepsilon_{j} ( \theta ) =\delta_{j,0}\, mR \, \cosh (\theta) +
\sum_{k=0}^{1} \int_{-\infty}^\infty \Phi_{j,k}(\theta-\theta') \log
\left( 1+ e^{-\varepsilon_k(\theta')}\, \right)\,\, d \theta'\ ,\qquad 
j=0,1. \label{eq:tbae1}
\end{equation}
As is well known, the free energy (\ref{eq:minfe1}) can be re-interpreted as 
the ground state energy 
\begin{equation}
E_{0}(R)= Rf_0(mR)={m^2 R\,\log(m R_0)\over8\pi}-\frac{m}{2 \pi}\int_{- 
\infty}^{\infty}\cosh\theta\,\log\left(\rule{0mm}{4mm}1+ e^{-\varepsilon_0
(\theta)}\,\right)\,\, d\theta      \label{eq:eofr1}
\end{equation}
of a finite-size system defined on a circle of circumference $R$. The 
leading asymptotics of $E_{0}(R)$ in the ultraviolet limit $R\rightarrow0$, 
can be calculated using the standard ``dilogarithm trick'' 
\cite{TsvelickWiegmann}. This results in
\beq
E_0(R) \sim - \frac{\pi}{10R} \label{eq:conf1}
\eeq
which is in agreement with the expected form
$E_0(R)\sim-(\pi/6R)(c\minus 24\Delta_{0})$, where $\Delta_{0}\! =\! -1/20$ 
is the lowest conformal dimension of the operator algebra associated with
the minimal CFT model $\mathcal{M}_{3,5}$.

\section{The dilute-$A_2$ lattice model}
\setcounter{equation}{0}

The IQFT model considered above is associated with a certain  
regime of the solvable dilute-$A_2$ lattice model 
\cite{WarnaarNienhuisSeaton,WarnaarPearceSeatonNienhuis}. The general 
dilute-$A_M$ model is an interaction-round-a-face (IRF) model defined on 
the square lattice which has discrete spin variables $a_i$ (or ``heights'')  
located at the lattice sites. These heights can take one of $M$ (with $M
\ge2$) possible integer values $1\le a_i\le M$, subject to the restriction 
that neighbouring sites either have the same height or differ by $\pm1$. 
The explicit expressions  for the Boltzmann weights of an elementary face 
of the lattice are given in Appendix~A. They are parameterised through 
the elliptic theta functions $\vartheta_j(u;p)$ of (spectral) variable 
$u$ and nome $p=e^{-\tau}$ which are also given in Appendix~A. Moreover, 
these Boltzmann weights depend on an additional  parameter  $\lambda$ which 
should be chosen from the discrete set of values
\beq
\lambda={k \pi \over 4 (M\plus 1)}, \qquad k\in \{1,\ldots,
M,M\plus 2,\ldots,2M\plus 1\} .   \label{lambda}
\eeq 
It is only when $k\! =\! M$ or $k\! =\! M\plus 2$ that the model is 
physical in the sense that all of the Boltzmann weights are real and 
positive (for some range of the variable $u$ on the real axis). Here we 
consider a specific unphysical case (corresponding to $k=1$ in 
(\ref{lambda})) with 
\beq 
M=2,\qquad p>0,\qquad 0<u<3\lambda,\qquad \lambda=\pi/12 . 
\label{eq:model}
\eeq
We show below that  the scaling limit of this model is described by the 
IQFT considered in the previous sections (\ie the IQFT obtained as a 
perturbation of the minimal non-unitary CFT ${\cal M}_{3,5}$ by its 
operator $\phi_{2,1}$). 

The row-to-row transfer matrix (with periodic boundary conditions in the
horizontal direction) is defined by
\beq
\bmath{T}_{\{a\}}^{\{b\}}(u)=\prod_{j=1}^N\boltzmannweight{b_{j}}{b_{j+1}}
{a_{j}}{a_{j+1}}{u}{}{} \label{eq:tmatrix2}
\eeq
where $\{a\}=\{a_1,a_2,\ldots,a_N\}$ and $\{b\}=\{b_1,b_2,\ldots,b_N\}$
denote heights on two consecutive rows of the lattice, periodic boundary
conditions imply that $a_{N+1}=a_1$ and $b_{N+1}=b_1$, and $N$ is the number 
of sites per row. Since the Boltzmann weights of the model satisfy the 
Yang-Baxter equation \cite{WarnaarNienhuisSeaton}, the transfer matrices 
$\bmath{T}(u)$ with different values of the spectral parameter $u$ commute 
among themselves (for fixed values of $p$ and $\lambda$)
\beq
[\bmath{T}(u),\ \bmath{T}(u')]=0
\eeq
and hence can be simultaneously diagonalized. 
The corresponding eigenvalues of the transfer 
matrix (\ref{eq:tmatrix2}) are given by \cite{BazhanovNienhuisWarnaar}
\bea
\Lambda(u) & = & \omega\left(\rule{0mm}{4mm}h(2\lambda\minus u)\;h(3\lambda
      \minus u)\right)^{N}\frac{Q(u+\lambda)}{Q(u\minus\lambda)} \nonumber \\ 
 & + & \left(\rule{0mm}{4mm}h(u)\; h(3\lambda\minus u)\right)^{N}\frac{Q(u)\; 
      Q(u\minus 3\lambda)}{Q(u\minus\lambda)\; Q(u\minus 2\lambda)},  
      \label{eq:eigval} \\
 & + & \omega^{-1}\left(\rule{0mm}{4mm}-h(u)\;h(\lambda\minus u)\right)^N 
      \frac{Q(u\minus 4\lambda)}{Q(u\minus 2\lambda)} \nonumber 
\eea
where
\beq
Q(u)=\prod_{j=1}^N h(u-i v_j)\label{Qmatr}
\eeq
and where the function $h(u)$ coincides (up to a simple factor) with the
standard theta function defined in (\ref{eq:thetafns})
\beq
h(u)=p^{-1/4}\;\te\left(u;p\right),\qquad p=e^{-\tau} .
\eeq
The numbers $\{v_j\}$ are solutions of the following set of Bethe Ansatz 
(BA) equations
\beq
\omega^{-1}\left(\frac{h(\lambda\plus iv_j)}{h(\lambda\minus iv_j)}
\right)^{N}=-\prod_{k=1}^{N}\frac{h(iv_j\minus iv_k\plus 2\lambda)\; h(i 
v_j \minus iv_k\minus\lambda)}{h(iv_j\minus iv_k\minus 2\lambda)\; h(iv_j 
\minus iv_k\plus\lambda)},\qquad j=1,\ldots,N   \label{BAE}
\eeq
with $\omega=\exp(i \pi m/(M\plus 1))$, $\; m=1,\ldots,M$.

Now consider the Hamiltonian of the associated one-dimensional quantum 
spin chain 
\beq
{\bf H}_N=-{1\over 4\delta}{d\over du} \log {\bf T}_N(u)\Big\vert_{u=0}
+{\rm const}\ ,\label{Hlat}
\eeq
where $\delta$ is a parameter with the dimension of length. The 
corresponding energy eigenvalues (with a suitable choice of the constant 
term in (\ref{Hlat})) can be expressed as  
\beq
E_N=-{1\over 4\delta}{d\over du} \sum_{j=1}^N\log\left.\left(\omega{h(u+
\lambda -iv_j)\over h(u-\lambda-iv_j)}\right)\right\vert_{u=0}\ .
\label{Elat}
\eeq
The scaling limit of this lattice model can also be analysed within the  
TBA approach. When $N\!\rightarrow\!\infty$ the solutions $\{v_j\}$ of 
(\ref{BAE}) converge to certain asymptotic patterns in the complex 
$v$-plane, which can be viewed as collections of ``strings'' 
\cite{TakahashiSuzuki}. An $\ell$-string is a set of $\ell\ge1$ complex 
roots with the same real part, symmetric with respect to the real axis of 
$v$ and equally spaced (with the spacing $2\lambda $) along the imaginary 
axis. The total number of roots  $\ell$ in a string  is called the length 
of the string. Numerical calculations suggest that for $N\rightarrow
\infty$ the string spectrum of (\ref{BAE}) is saturated by the strings of 
lengths $\ell=1,2,3$ only (in the sense that the number of strings of 
other types in a generic solution of (\ref{BAE}) grows slower than $N$).
Assuming this picture is correct, it is possible to write the roots $v_j$ 
which solve (\ref{BAE}) in the form
\beq
v_{j,k}^{(\ell)}={1\over4}\theta_{j}^{(\ell)}+i\Delta^{(\ell)}_k+O(e^{-a
N}),\qquad k=1,\ldots ,\ell,\quad j=1,\ldots,N^{(\ell)},\qquad\ell=1,2,3
\label{strings}
\eeq
where 
\beq
\Delta^{(\ell)}_k=\lambda(2k\minus\ell\minus 1),
\eeq
and the (real) numbers $\theta_{j}^{(\ell)}$ determine the centres of the 
strings. It is to be noted that for non-vanishing values of the elliptic 
nome $p>0$ 
the convergence parameter $a$ in (\ref{strings}) is positive and non-zero. 
The integers $N^{(\ell)}$ denote the total number of strings of length  
$\ell$ in a particular solution of (\ref{BAE}). As follows from 
(\ref{Qmatr}) these numbers are restricted by
\beq
\sum_{\ell=1}^{3}\ell N^{(\ell)}=N+o(N).\label{eq:srest}
\eeq

In the thermodynamic limit $N\to\infty$, the centres of the strings form 
continuous distributions and the BA equations (\ref{BAE}) lead to the
following integral
equations\footnote{\label{foot} This approximation is valid provided $N
p^2\gg1$ in addition to $N\gg1$. In this section we will assume that this 
extra condition is also satisfied.} for the densities of strings 
$\rho_\ell(\theta)$ and ``holes'' $\tilde\rho_\ell(\theta)$  \cite{Yang2}
\beq
b_{s}(\theta )=\widetilde{\rho}_{s}(\theta)+{\rho}_{s}(\theta)+\sum_{t=1
}^{3}\int_{-2\tau}^{2\tau}B_{s,t}(\theta\minus\theta^{\prime})\rho_{t}(
\theta^{\prime})d\theta^{\prime}, \qquad s,t=1,\ldots ,3\ ,\label{BAEL}
\eeq
where 
\beq
b_s(\theta)=\sum_{k=1}^{s}a_2\left(\theta+i\Delta^{(s)}_k\right)
\eeq
\beq
B_{s,t}(\theta )=\sum_{k=1}^{s}\sum_{m=1}^{t}\left[ a_4\left(\theta+i
\Delta^{(s)}_k+i\Delta^{(t)}_m\right)-a_2\left(\theta+i\Delta^{(s)}_k 
+i\Delta^{(t)}_m\right) \right] 
\eeq
and where the function $a_{j}(\theta)$ is defined as 
\beq
a_{j}(\theta )=\frac{1}{2\pi i}\frac{d}{d\theta}\log\left[\frac{h(\lambda 
j/2+i\theta/4)}{h(\lambda j/2-i\theta/4)}\right] .
\eeq
Note that the functions $a_j(\theta)$ (as well as $b_s(\theta)$, $B_{s,t}
(\theta)$ and all of the densities $\rho_\ell(\theta)$ and $\tilde
\rho_\ell(\theta)$)  are periodic with period $4\tau$. Also, the 
densities $\rho_\ell$ are normalised by the condition
\beq
\int_{-2\tau}^{2\tau} \rho_{\ell}(\theta )d\theta =N^{(\ell )}/N
\eeq
which from (\ref{eq:srest}) implies that
\beq
\sum_{\ell=1}^{3}\int_{-2\tau}^{2\tau}\ell\rho_{s}(\theta )d\theta =1 .
\eeq
This relation together with equation (\ref{BAEL}) (with $s=3$) implies 
that $\tilde\rho_3(\theta)\equiv0$, and hence it is possible to eliminate
the density $\rho_3(\theta)$ from (\ref{BAEL}). Re-denoting the remaining 
densities as
\beq
\begin{array}{ll} \rho_1(\theta)\equiv\delta\,\tilde\sigma_0(\theta\plus 
2\tau),\qquad & \tilde\rho_1(\theta)\equiv\delta\,\sigma_0(\theta\plus 2
\tau)\qquad \\ \rho_2(\theta)\equiv\delta\,\sigma_1(\theta\plus 2\tau),
\qquad & \tilde\rho_2(\theta)\equiv\delta\,\tilde\sigma_1(\theta\plus 2
\tau)\qquad\end{array}
\eeq
where the parameter $\delta$ is the same as in (\ref{Hlat}), one obtains 
the integral equations
\beq
\delta^{-1}\delta_{j,0}\,\phi_0(\theta\plus 2\tau,\tau)=\sigma_j(\theta)+
\widetilde{\sigma}_j(\theta)+\sum_{k=0}^1\int_{-\infty}^\infty\Phi_{j,k}
(\theta\minus\theta',\tau)\,\,\sigma_k(\theta')\,\, d\theta', \qquad j=0,1. 
\label{BA}
\eeq
Here, the functions 
\beq
\Phi_{j,k}(\theta,\tau)=(\delta_{j,k}-\delta_{j,k-1}-\delta_{j,k+1})
\phi_0(\theta,\tau)   \label{Phitau}
\eeq
{}
\beq
\phi_0(\theta,\tau)={1\over4\tau}\sum_{n=-\infty}^{\infty}e^{i\theta x_n}
{\cosh(2\lambda x_n)\over\cosh(6\lambda x_n)},\qquad x_n={\pi n\over 2\tau}, 
\label{phitau} 
\eeq      
and the parameter $\lambda={\pi/12}$. The largest eigenvalue $\Lambda_0(u)$ 
of the transfer matrix (\ref{eq:tmatrix2}) (and hence the ground state 
energy of the Hamiltonian (\ref{Hlat})) is determined by the solution of  
(\ref{BAE}) that consists only of $1$-strings. The corresponding density 
\beq
\rho_1^{(vac)}(\t)=\phi_0(\t,\tau)
\eeq
can be easily determined from (\ref{BA}). Then from (\ref{eq:eigval}) one 
obtains
\beq
\log\kappa(u)=\lim_{N\to \infty}\left({1\over N}\,\log\Lambda_0(u)\right)=
\log\rho +\!
\int_{-2\tau}^{2\tau}\!\!\log\left({\vartheta_1(u\plus\lambda\plus i\theta/4)
\vartheta_1(4\lambda\minus u\plus i\theta/4)\over\vartheta_1(\lambda\plus
i\theta/4)\vartheta_1(4\lambda\plus i\theta/4)}\right)\phi_0(\theta,\tau)
d\theta    \label{PF}
\eeq
where 
\beq
\rho = h(2\lambda )h(3\lambda ) .
\eeq
Finally, the spectrum of the Hamiltonian (\ref{Elat}) for large $N$ reads 
\beq
E_N=-{N\over 4\delta} {d\over du} \log \kappa(u)\Big\vert_{u=0}+2\pi N
\int_{-2\tau}^{2\tau}\phi_0(\theta\plus 2\tau,\tau)\,\sigma_0\,(\theta)d\theta
\label{Elat1}
\eeq
where the first term represents the vacuum energy and the second term 
represents the excitation energy. The gap in the excitation 
spectrum is then determined by the value of $\phi_0(\theta\plus 2\tau,\tau)$ 
at $\theta=0$ where the one-particle excitation energy has a minimum.

Now consider the scaling limit. The leading singularity of (\ref{PF})
for $p\rightarrow 0$ is given by
\beq
\log \kappa(u)_{\rm sing}=-{12 \sin \,4u\over\pi}\,p^4 \log p 
\label{ksing}
\eeq
which determines the thermal exponent $\alpha$ and correlation length
exponent $\nu$ to be 
\beq
2-\alpha=4,\qquad \nu=2 .
\eeq
Let us now identify the parameter $\delta$ in (\ref{Hlat}) with the 
lattice spacing constant, and take the limits $N\rightarrow\infty$ and
$\delta\rightarrow0$, while keeping both the (dimensional) length of the 
chain $L=N\delta$ and the correlation length $R_c\sim p^{-\nu}\delta$ 
finite. This means that
\beq
p^2\sim N^{-1},\qquad\delta \sim N^{-1},\qquad N\to \infty.
\label{slimit}
\eeq
We shall also assume that $L\gg R_c$ (which is equivalent to $Np^2\gg1$, 
see footnote 1 on page 8). In the above limit, the gap $m$ in 
the one-particle spectrum of (\ref{Elat1}) (\ie the mass gap) tends to 
a finite value. Indeed, one has
\beq
2\pi\delta^{-1}\phi_0(\theta\plus 2\tau,\tau)=m \cosh\theta
+O(p^2)\label{philim},\qquad p\rightarrow 0,
\eeq
where 
\beq
m=4\sqrt{3}p^2\,\delta^{-1},
\eeq
and hence expression (\ref{Elat1}) becomes 
\beq
E(L)={m^2\,L\log(mL_0)\over8\pi}+mL\int_{-\infty}^{\infty}\cosh\theta\, 
\sigma_0\,d\theta .    \label{Elat2}
\eeq
It is to be noted that the bulk vacuum energy term here is determined
entirely by the 
singular part (\ref{ksing}) of the largest eigenvalue of the transfer matrix. 
The parameter $L_0$ is non-universal, and in particular, it depends on an 
overall shift of the energy spectrum in (\ref{Hlat}). It is obvious that
the energy expression (\ref{Elat2}) is identical to that given by 
(\ref{Eqft}) and (\ref{Epsqft}) of the previous section. 
Similarly, using (\ref{philim}) and 
\beq
\phi_0(\theta,\tau)=\phi(\theta)+O(p^2),\qquad  p\rightarrow 0
\label{philim1}
\eeq
with $\phi(\theta)$ given by (\ref{phi}), it is easy to see that  
the integral equations (\ref{BA}) in the limit (\ref{slimit}) become
identical to (\ref{eq6:denseq1}). And finally, the density of states in 
this lattice model is, of course, determined by exactly the same 
combinatorial entropy functional as in (\ref{eq6:entropy}).

\section{Excited states energies for the finite-size system}
\setcounter{equation}{0}

In the previous section we established an equivalence between the QFT  
(\ref{action}) and the scaling limit of the dilute $A_2$ lattice model. 
We have shown that the density of states in these two systems are 
described by the same integral equations. This means, in particular, 
that they also have the same finite-volume ground state energy (as 
determined by the TBA approach). It is natural to expect that this 
correspondence extends to all (finite-volume) excitation energies.  

An effective way to extract information about the spectrum of the 
transfer matrix in lattice models is to use the 
functional relations. Using the standard fusion procedure,
it is not difficult to show by explicit calculations\footnote{See also
Sect.5.3 below.} that in the case of the 
dilute-A$_2$ model considered here, the transfer matrix
(\ref{eq:tmatrix2}) 
(and hence all of its eigenvalues) satisfies the relation
\beq
\bmath{T}(u\plus\pi/12)\bmath{T}(u\minus\pi/12)=a(u)\bmath{T}(u
\plus\pi/2)+(-1)^{N}\,b(u)\bmath{T}(u) \label{eq:fusionreln}
\eeq
where $N$ is the number of sites per row (which hereafter is assumed 
to be even), and the scalar coefficients read
\beq
%\begin{array}{l} 
a(u)=\left(h\left(u - \pi/12
\right)h\left(u - \pi/6\right)\right)^{N},\qquad
%\\
b(u)=\left(h\left(u +\pi/12\right)h\left(u
-\pi/3\right)\right)^{N}\ .
%\end{array} 
\eeq
 
In references \cite{KlumperPearce1,KlumperPearce2}, Kl\"umper and Pearce 
developed a technique 
for transforming functional relations of the form (\ref{eq:fusionreln}) to 
integral equations. This is particularly useful in studying the eigenvalues 
of the transfer matrix for finite values of $N$. To apply this technique 
one first requires information on the patterns of zeroes of the 
eigenvalues. We have used numerical calculations to study these patterns as
follows.
 
For small system sizes all solutions to the Bethe-Ansatz  equations 
(\ref{BAE}) are easily found numerically using standard nonlinear 
equation solving algorithms. The eigenvalue expressions (\ref{eq:eigval}) 
corresponding to these BA solutions, can then be compared against the results 
of the direct numerical diagonalization of the transfer matrix 
(\ref{eq:tmatrix2}). This enables one to numerically determine the patterns 
of zeroes of $\Lambda (u)$ (or equivalently, patterns of zeroes of $Q(u)$) 
corresponding to each eigenvalue of the transfer matrix (\ref{eq:tmatrix2}).
The results obtained for the largest and a few next-to-largest eigenvalues 
of the transfer matrix $\bmath{T}(u)$ are presented below.
In the following, it will be convenient to use a new spectral variable
$\theta$, which is 
related to the variable $u$ by
\beq
\theta=-4i(u\minus\pi/8\minus i\tau/2) .  \label{theta-u}
\eeq
To make this change more explicit we denote the eigenvalues of the 
transfer matrix as $T(\theta)\equiv\Lambda(u)$ (assuming that $\theta$ 
and $u$ are always related by (\ref{theta-u})). The eigenvalues $T(
\theta)$ are entire functions of $\theta$ satisfying the (quasi-) 
periodicity relations 
\beq
T(\theta\plus 4\pi i)=T(\theta),\qquad T(\theta\plus 4\tau)= \exp\left[ 
N(\theta\plus 4\tau )\right] T(\theta). \label{eq:quasipp}
\eeq
In fact, these eigenvalues can be written as the products 
\beq	
T(\theta)=C\  e^{\alpha \theta} \ \prod_{j=1}^{2N}
\vartheta_1\Big(i(\theta\minus\theta_j)/4\Big) \label{eq:products}
\eeq
where $C$ and $\alpha = \sum_{j=1}^{2N}\theta_{j}$ are some constants 
and $\theta_j$ denote the zeroes of $T(\theta)$. The patterns of these 
zeroes for the three
largest eigenvalues (which we denote as $T^{(0)}(\theta)$,\ \ $T^{(1)}(\theta)$
\ and \ $T^{(2)}(\theta)$) with $p =0.1$, obtained numerically for $N=14$
sites per row are shown in Figures 1, 2 and 3a respectively. The
zeroes of $T^{(2)}(\theta)$ with $p =0.3$ are also shown in Figure 3b
to emphasise the qualitative change in the position of two of the
zeroes at some intermediate value of $p$. The total number of zeros
in each of the figures is equal to $2N=28$.
\begin{figure}[t]
\begin{center}
\includegraphics[scale=0.5,angle=-90]{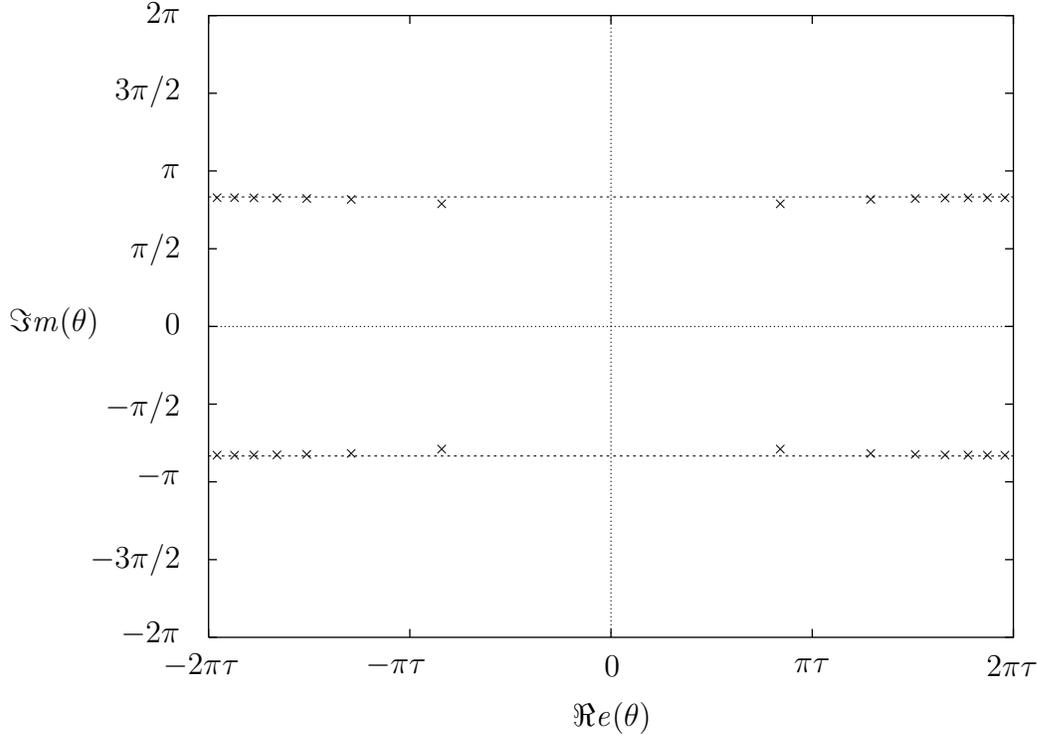}
   {\setlength{\unitlength}{7mm}
   \begin{picture}(0,-16.0)(0,0)
   \put(-16.8,-0.3){\makebox(0,0)[r]{\mbox{$2\pi$}}}
   \put(-16.8,-1.8){\makebox(0,0)[r]{\mbox{$3\pi/2$}}}
   \put(-16.8,-3.3){\makebox(0,0)[r]{\mbox{$\pi$}}}
   \put(-16.8,-4.8){\makebox(0,0)[r]{\mbox{$\pi/2$}}}
   \put(-16.8,-6.2){\makebox(0,0)[r]{\mbox{$0$}}}
   \put(-16.8,-7.8){\makebox(0,0)[r]{\mbox{$-\pi/2$}}}
   \put(-16.8,-9.1){\makebox(0,0)[r]{\mbox{$-\pi$}}}
   \put(-16.8,-10.7){\makebox(0,0)[r]{\mbox{$-3\pi/2$}}}
   \put(-16.8,-12.1){\makebox(0,0)[r]{\mbox{$-2\pi$}}}
   \put(-16.4,-12.7){\makebox(0,0)[c]{\mbox{$-2\pi\tau$}}}
   \put(-12.7,-12.7){\makebox(0,0)[c]{\mbox{$-\pi\tau$}}}
   \put(-8.6,-12.7){\makebox(0,0)[c]{\mbox{$0$}}}
   \put(-4.8,-12.7){\makebox(0,0)[c]{\mbox{$\pi\tau$}}}
   \put(-1.0,-12.7){\makebox(0,0)[c]{\mbox{$2\pi\tau$}}}
   \put(-8.6,-13.7){\makebox(0,0)[c]{\mbox{$\real (\theta )$}}}
   \put(-19.2,-6.2){\makebox(0,0)[c]{\mbox{$\imag (\theta )$}}}
   \end{picture}}
   \vspace{5mm}
\end{center}
\caption{Zeroes of the largest eigenvalue $T_{0}(\theta )$ 
for $N=14$ and $\p =0.1$.}
\end{figure}
\begin{figure}[ht]
\begin{center}
\includegraphics[scale=0.5,angle=-90]{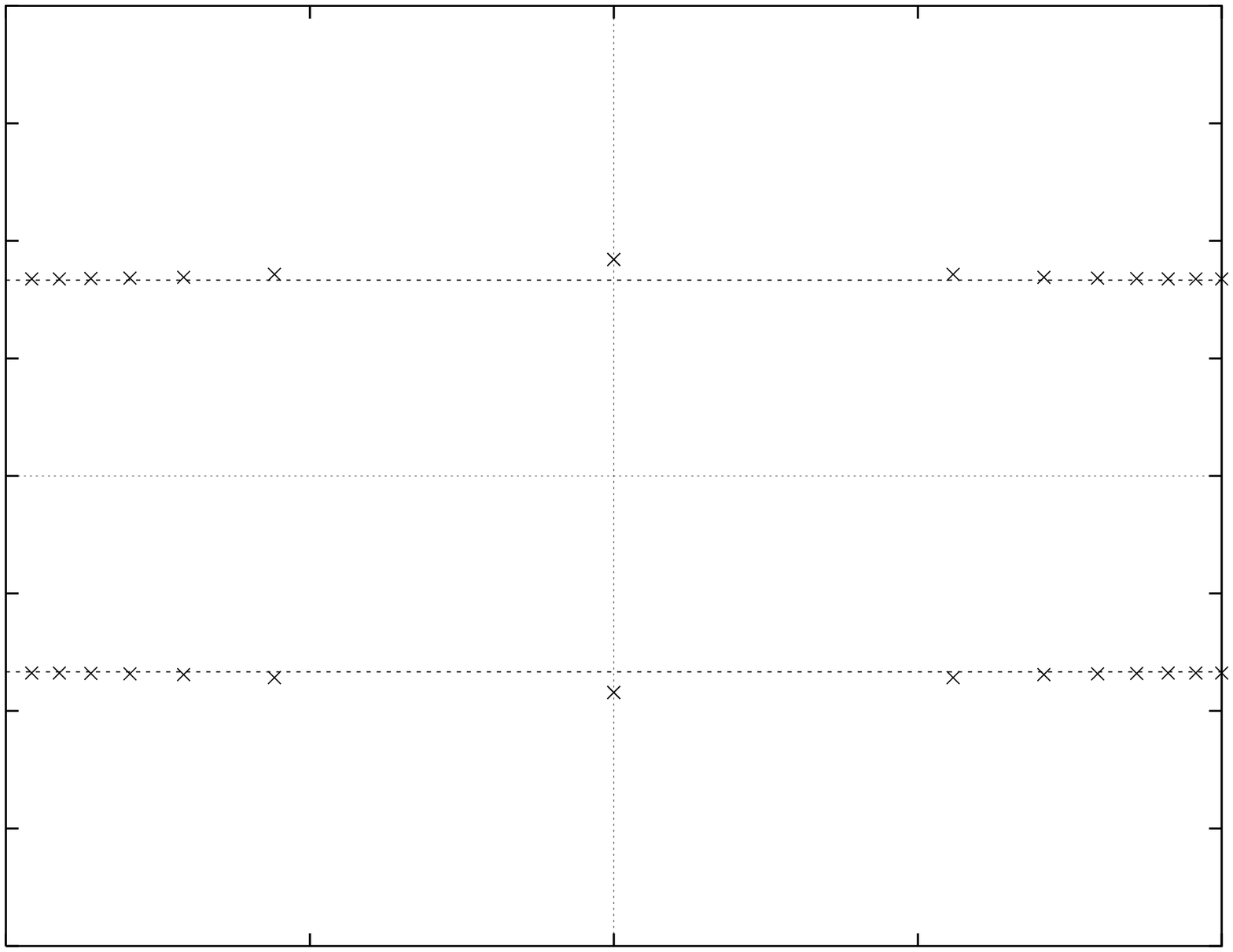}
   {\setlength{\unitlength}{7mm}
   \begin{picture}(0,-16.0)(0,0)
   \put(-16.8,-0.3){\makebox(0,0)[r]{\mbox{$2\pi$}}}
   \put(-16.8,-1.8){\makebox(0,0)[r]{\mbox{$3\pi/2$}}}
   \put(-16.8,-3.3){\makebox(0,0)[r]{\mbox{$\pi$}}}
   \put(-16.8,-4.8){\makebox(0,0)[r]{\mbox{$\pi/2$}}}
   \put(-16.8,-6.2){\makebox(0,0)[r]{\mbox{$0$}}}
   \put(-16.8,-7.8){\makebox(0,0)[r]{\mbox{$-\pi/2$}}}
   \put(-16.8,-9.1){\makebox(0,0)[r]{\mbox{$-\pi$}}}
   \put(-16.8,-10.7){\makebox(0,0)[r]{\mbox{$-3\pi/2$}}}
   \put(-16.8,-12.1){\makebox(0,0)[r]{\mbox{$-2\pi$}}}
   \put(-16.4,-12.7){\makebox(0,0)[c]{\mbox{$-2\pi\tau$}}}
   \put(-12.7,-12.7){\makebox(0,0)[c]{\mbox{$-\pi\tau$}}}
   \put(-8.6,-12.7){\makebox(0,0)[c]{\mbox{$0$}}}
   \put(-4.8,-12.7){\makebox(0,0)[c]{\mbox{$\pi\tau$}}}
   \put(-1.0,-12.7){\makebox(0,0)[c]{\mbox{$2\pi\tau$}}}
   \put(-8.6,-13.7){\makebox(0,0)[c]{\mbox{$\real (\theta )$}}}
   \put(-19.2,-6.2){\makebox(0,0)[c]{\mbox{$\imag (\theta )$}}}
   \end{picture}}
   \vspace{5mm}
\end{center}
\caption{Zeroes of the 2nd largest eigenvalue $T_{1}(\theta )$ 
for $N=14$ and $\p =0.1$.}
\end{figure}
\begin{figure}[p]
\begin{center}
\vspace{-5mm}
\includegraphics[scale=0.5,angle=-90]{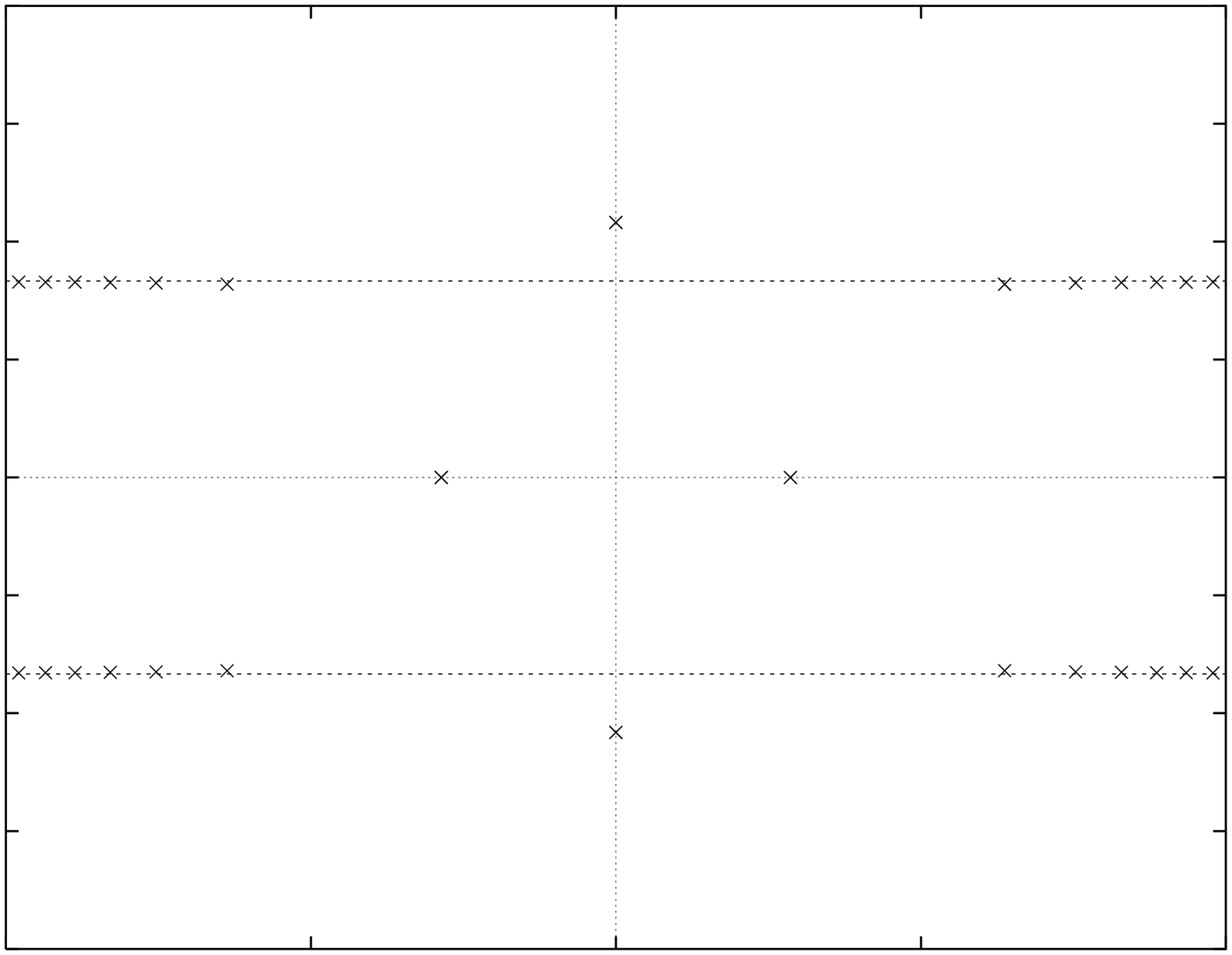}
   {\setlength{\unitlength}{7mm}
   \begin{picture}(0,-16.0)(0,0)
   \put(-16.8,-0.3){\makebox(0,0)[r]{\mbox{$2\pi$}}}
   \put(-16.8,-1.8){\makebox(0,0)[r]{\mbox{$3\pi/2$}}}
   \put(-16.8,-3.3){\makebox(0,0)[r]{\mbox{$\pi$}}}
   \put(-16.8,-4.8){\makebox(0,0)[r]{\mbox{$\pi/2$}}}
   \put(-16.8,-6.2){\makebox(0,0)[r]{\mbox{$0$}}}
   \put(-16.8,-7.8){\makebox(0,0)[r]{\mbox{$-\pi/2$}}}
   \put(-16.8,-9.1){\makebox(0,0)[r]{\mbox{$-\pi$}}}
   \put(-16.8,-10.7){\makebox(0,0)[r]{\mbox{$-3\pi/2$}}}
   \put(-16.8,-12.1){\makebox(0,0)[r]{\mbox{$-2\pi$}}}
   \put(-16.4,-12.7){\makebox(0,0)[c]{\mbox{$-2\pi\tau$}}}
   \put(-12.7,-12.7){\makebox(0,0)[c]{\mbox{$-\pi\tau$}}}
   \put(-8.6,-12.7){\makebox(0,0)[c]{\mbox{$0$}}}
   \put(-4.8,-12.7){\makebox(0,0)[c]{\mbox{$\pi\tau$}}}
   \put(-1.0,-12.7){\makebox(0,0)[c]{\mbox{$2\pi\tau$}}}
   \put(-8.6,-13.5){\makebox(0,0)[c]{\mbox{$\real (\theta )$}}}
   \put(-19.2,-6.2){\makebox(0,0)[c]{\mbox{$\imag (\theta )$}}}
   \put(-8.6,-14.5){\makebox(0,0)[c]{\mbox{(a)}}}
   \end{picture}}
   \vspace{20mm}

\includegraphics[scale=0.5,angle=-90]{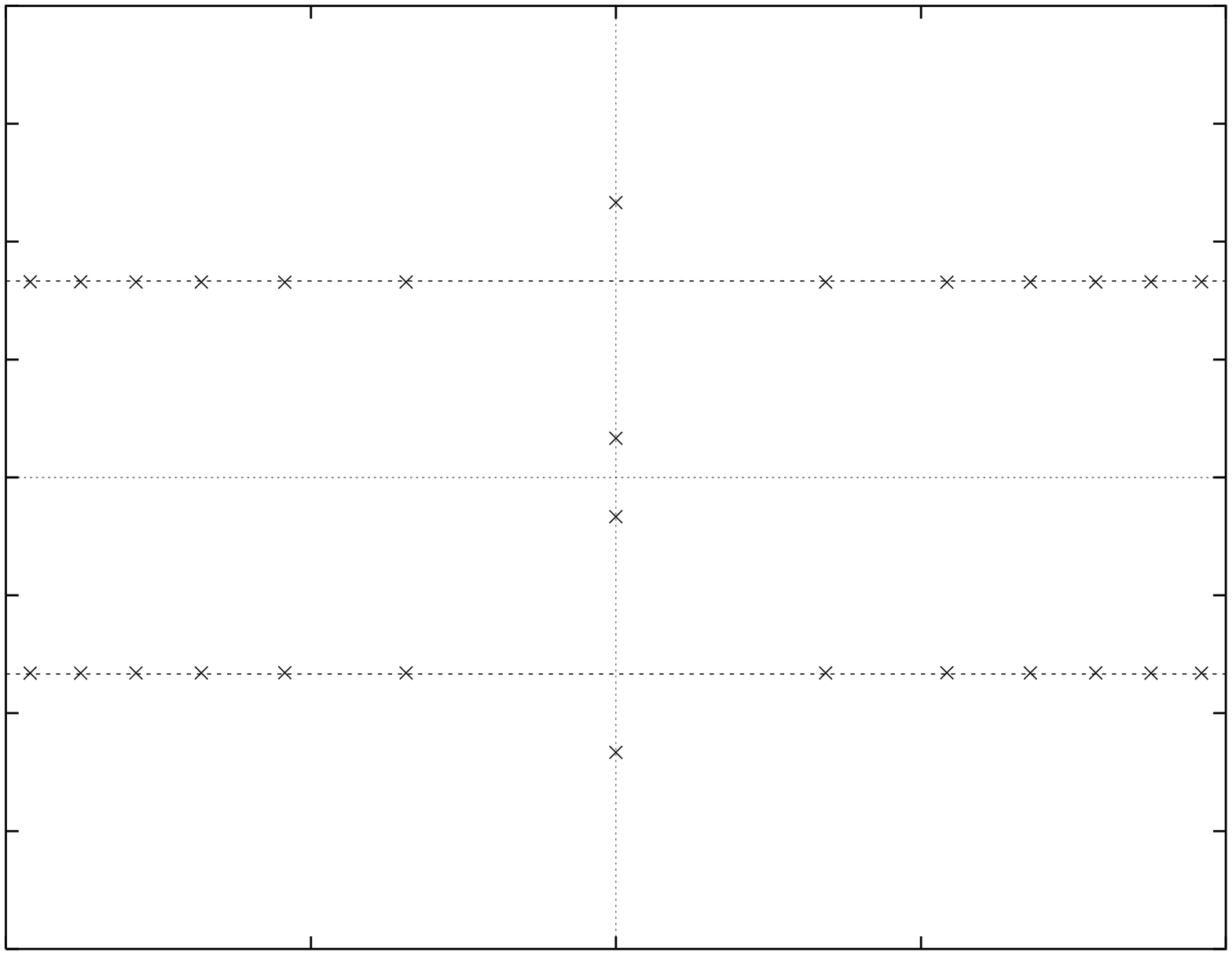}
   {\setlength{\unitlength}{7mm}
   \begin{picture}(0,-16.0)(0,0)
   \put(-16.8,-0.3){\makebox(0,0)[r]{\mbox{$2\pi$}}}
   \put(-16.8,-1.8){\makebox(0,0)[r]{\mbox{$3\pi/2$}}}
   \put(-16.8,-3.3){\makebox(0,0)[r]{\mbox{$\pi$}}}
   \put(-16.8,-4.8){\makebox(0,0)[r]{\mbox{$\pi/2$}}}
   \put(-16.8,-6.2){\makebox(0,0)[r]{\mbox{$0$}}}
   \put(-16.8,-7.8){\makebox(0,0)[r]{\mbox{$-\pi/2$}}}
   \put(-16.8,-9.1){\makebox(0,0)[r]{\mbox{$-\pi$}}}
   \put(-16.8,-10.7){\makebox(0,0)[r]{\mbox{$-3\pi/2$}}}
   \put(-16.8,-12.1){\makebox(0,0)[r]{\mbox{$-2\pi$}}}
   \put(-16.4,-12.7){\makebox(0,0)[c]{\mbox{$-2\pi\tau$}}}
   \put(-12.7,-12.7){\makebox(0,0)[c]{\mbox{$-\pi\tau$}}}
   \put(-8.6,-12.7){\makebox(0,0)[c]{\mbox{$0$}}}
   \put(-4.8,-12.7){\makebox(0,0)[c]{\mbox{$\pi\tau$}}}
   \put(-1.0,-12.7){\makebox(0,0)[c]{\mbox{$2\pi\tau$}}}
   \put(-8.6,-13.5){\makebox(0,0)[c]{\mbox{$\real (\theta )$}}}
   \put(-19.2,-6.2){\makebox(0,0)[c]{\mbox{$\imag (\theta )$}}}
   \put(-8.6,-14.5){\makebox(0,0)[c]{\mbox{(b)}}}
   \end{picture}}
   \vspace{10mm}

\end{center}
\caption{(a) Zeroes of the 3rd largest eigenvalue $T_{2}(\theta )$ 
for $N=14$ and $\p =0.1$, which corresponds to $r<r_{c}$;\mbox{\hspace{2mm}}
(b) Zeroes of the 3rd largest eigenvalue $T_{2}(\theta )$ for $N=14$ and $\p
=0.3$, which corresponds to $r>r_{c}$.}
\end{figure}
As can be seen from the figures, most of the zeroes accumulate  
near the lines 
\beq
\imag \theta= \pm{5\pi\over6} .
\eeq
In fact, for the largest eigenvalue $T^{(0)}(\theta)$ all of the zeroes are 
located near these lines. The eigenvalue of the first excited state 
$T^{(1)}(\theta)$ has two zeroes shifted to the positions
\beq
\theta=\pm i\pi+O(p) . \label{eq:szer}
\eeq
The eigenvalue of the second excited state $T_{2}(\theta )$ also has two 
shifted zeroes similar to (\ref{eq:szer}), and in addition, has two
more shifted zeroes inside the strip $|\imag\theta |<\pi/6$ which are
located at
\beq
\theta=\pm\gamma .   \label{eq:twozeroes} 
\eeq
The value of $\gamma $ is controlled by the dimensionless scaling 
parameter 
\beq
r=4\sqrt{3}\, p^2\, N. \label{eq:scalpar}
\eeq
Numerical calculations show that $\gamma$ vanishes when $r$ approaches 
some critical value\footnote{Note, that an accurate value of this
critical value $r_{c}$ 
is quite difficult to obtain due to the singular nature of this
point which leads to severe numerical instabilities.} $r_c$
\beq
\gamma=0, \qquad {\rm for }\qquad r=r_c\approx 2.85\pm 0.10   
\label{eq6:critr}
\eeq
For $r>r_c$ the parameter $\gamma$ is purely imaginary 
and approaches the value $\gamma_\infty={i\pi/6}$ as $r\to\infty$.
For $r<r_c$ the parameter $\gamma$ is real and increases as $O(|\log r|)$
when $r\to 0$. We found also  that the ratio $T(\t)/T(\t\plus 2\pi i)$ 
satisfies the condition
\beq
T(\theta)/T(\theta\plus 2\pi i)\in\reals, \qquad \mbox{ for }\imag (\theta
)=0.    \label{cond}
\eeq
In fact, this ratio is positive for $T^{(0)}(\theta)$, and for
$T^{(2)}(\theta)$ when 
$r>r_c$, and is negative for $T^{(1)}(\theta)$. For $T^{(2)}(\theta)$
with $r<r_c$, 
it is negative between the zeroes in (\ref{eq:twozeroes}) (\ie for $|\theta|
<\gamma$). Here, we assume that these properties of $T^{(0)}(\theta)$,\
$T^{(1)}(\theta)$ and $T^{(2)}(\theta)$ remain applicable for all arbitrarily large
values of $N$, and small $p$.

For the following calculations it is convenient to pass directly to the 
scaling limit (\ref{slimit}), defining the mass parameter $m$ and the 
system size $R$ as
\beq
m=4\sqrt{3}\, p^2\,\delta^{-1},\qquad\qquad R=N\delta  \label{size}
\eeq
where $\delta$ is the dimensional lattice spacing parameter. In this limit,
the appropriately normalised eigenvalues of the transfer matrix tend to 
the finite values
\beq
\mathbb{T}(\theta)=e^{\frac{r}{2\sqrt{3}}}\lim_{N\to\infty}\left[\exp\left(
c_1 N^{1\over2}\cosh(\theta/2)+c_2\log N\cosh\theta\right)\ T(\theta)
\right]   \label{eq:bigt}
\eeq
where the exponential multiplier in front of the limit sign is introduced 
for later convenience. The constants $c_1$ and $c_2$ should be chosen to 
regularize the product over zeroes in (\ref{eq:products}) which diverges 
in this limit. Estimating the asymptotic density of zeros $\theta_j$ for 
large $\theta$ from the Bethe Ansatz equations (\ref{BAE}), one can show 
that
\beq 
c_1=\frac{r^{\frac{1}{2}}}{3^{\frac{1}{4}}\sqrt{2}} 
\eeq
where $r\! =\! mR$ is the dimensionless scaling parameter previously defined 
in (\ref{eq:scalpar}). Although it is also possible to calculate the value of
the constant $c_2$, this is not essential for the following discussion and is
hence omitted here. It can be shown from the properties (\ref{eq:quasipp}),
that the functions $\mathbb{T}(\theta)$ defined in (\ref{eq:bigt}), are 
entire functions of the variable $\theta$ which obey the periodicity
relation
\beq
\mathbb{T}(\theta\plus 4\pi i)=\mathbb{T}(\theta)
\eeq
and have the following asymptotic behaviour for  $\theta\to\pm\infty$
\beq
\begin{array}{l}
\log\left|\rule{0mm}{4mm}\mathbb{T}(\theta)\right|\sim \mp\frac{\sqrt{3}\,
r}{8\pi}\theta e^{\pm\theta}+c_{3}e^{\pm\theta}+O(e^{\mp \theta}) \\
\log\left|\rule{0mm}{4mm}\mathbb{T}(\theta)/\mathbb{T}(\theta\plus 2\pi 
i)\right|
\sim{r\over2}\, e^{\pm\theta}+O(e^{\mp\theta}) \rule{0mm}{6mm}
\end{array}    \label{eq:bigtasymp}
\eeq
where $c_3$ is some (unknown) constant.
These asymptotics are valid in the strip\footnote{In fact, these 
asymptotics are valid in a wider strip $|\imag\theta|<5\pi/6-\epsilon$
for arbitrary small positive $\epsilon$.} $|\imag\theta|<\pi/3$. 
The functions $\mathbb{T}(\theta )$ are in fact the QFT counterparts of 
the eigenvalues of the transfer matrix of the lattice model. Using 
(\ref{eq:fusionreln}) and (\ref{eq:bigt}) it is easy to see that they 
satisfy the functional relation
\beq
\mathbb{T}(\theta\plus i\pi/3)\mathbb{T}(\theta\minus i\pi/3)=e^{{r
\over4}\cosh\t}\ \mathbb{T}(\theta)+e^{-{r\over4}\cosh\t}\ \mathbb{T}(
\theta+2\pi i) .  \label{Fnrel}
\eeq
Introducing the new functions 
\beq
\mathbb{Y}_0(\t)={e^{{1\over2}r\cosh\t}\  \mathbb{T}(\t)/\mathbb{T}
(\t\plus 2\pi i)},\qquad \mathbb{Y}_1(\t)=\mathbb{Y}_0(\t\plus 2\pi i)
\label{eq:newfunc}
\eeq 
one can then rewrite the functional relation (\ref{Fnrel}) in the form
\beq
{\mathbb{Y}_{0} (\theta\plus i\pi/3) \mathbb{Y}_{0} (\theta\minus 
i\pi/3)\over \mathbb{Y}_{0} (\theta)} = \frac{1+\mathbb{Y}^{-1}_{0}(
\theta)}{1+ \mathbb{Y}_{1}^{-1} (\theta)} .    \label{eq:newrel}
\eeq
The asymptotic behaviour of the functions $\mathbb{Y}_{0,1}(\t)$ at  
$\t\to\pm\infty$ is determined by (\ref{eq:bigtasymp}) to be
\beq
\log  |\mathbb{Y}_{0}(\t)|={r\over2}\,e^{\pm\t} + O(e^{\mp\t}), \qquad
\log  |\mathbb{Y}_{1}(\t)|=O(e^{\mp\t}), \qquad |\imag \t|<\pi/3 .    
\label{eq:newasymp}
\eeq

The derivation of  the Kl\"umper-Pearce equations for the finite-size energies
of the QFT (\ref{action}), is based on the following simple fact 
\cite{BazhanovLukyanovZamolodchikov4}
that if $f(\theta)$ is a function which is regular,
bounded in the strip $|\imag (\theta )|<\pi/3 $, and satisfies the
relation
\beq
f(\theta\plus i\pi/3)+f(\theta\minus i\pi/3)-f(\theta) = g(\theta)\label{fact}
\eeq
for some function $g(\theta)$, then
\beq
f(\theta )=\int_{-\infty}^{\infty}d\theta'\ \phi (\theta\minus\theta')\,
g(\theta')\ ,\label{fact1}
\eeq
where $\phi(\theta)$ is defined in (\ref{phi}).

The eqs.(\ref{fact}),(\ref{fact1}) combined with knowledge of the analytic
properties and asymptotic behaviour of the functions $\mathbb{Y}_{0,1}(\t)$ 
defined in (\ref{eq:newrel}), allows one to rewrite the functional equation 
(\ref{eq:newrel}) in terms of nonlinear integral equations which
generalise the TBA equations. From the functional equation (\ref{eq:newrel}),
it is easy to see that the simplest case where $\mathbb{Y}_{0}(\theta)$ and
$\mathbb{Y}_{1}(\theta )$ have no zeroes in the strip $\imag (\theta
)\in (-\pi /3,\pi /3)$, only requires a straightforward application
of the lemma (see Section 4.1 below for further details).
From Figures 1, 2, 3a and 3b, it can be seen that the only complication
arising for the next few largest eigenvalues involves the function
$\mathbb{T}(\theta)$ having a finite number of zeroes in the strip $|
\imag\theta |<\pi/3$, which are either real or occur in complex
conjugated pairs. Suppose now, for generality, that the function $\mathbb{T}
(\theta)$ has a total of $A+2B$ zeroes in this strip where there are $A$
real zeroes $\alpha_a$, $a=1,\ldots,A$, and $B$ complex conjugated
pairs $\beta_b\pm i\gamma_b$, $b=1,\ldots,B$. Then using (\ref{Fnrel})
and (\ref{eq:newfunc}) it is easy to show that 
the eigenvalue expression (\ref{Elat}) in the scaling limit
(\ref{slimit}) can then be written as
\beq
\begin{array}{c} \displaystyle E(R)={m^2 R\,\log(m R_0)\over8\pi}+{
\sqrt{3}m\over2}\sum_{a=1}^A e^{\alpha_a}+2m\sum_{b=1}^B\cosh\beta_b\,
\sin\left({\pi\over3}-\gamma_b\right)\\ \displaystyle -\frac{m}{2\pi}
\int_{-\infty}^{\infty}\cosh\theta\,\log\left|\rule{0mm}{4mm}1+\mathbb{Y}_0
(\theta)^{-1}\,\right|\,\, d\theta . \\ \label{eq:eofr2} \end{array}
\eeq
In Sections 4.1-4.3 below, we explicitly study the form of the resulting
TBA integral equations and energies corresponding to the ground state and
first two excited state energies of the spectrum (\ref{Elat}).

\subsection{The TBA Equations for the Ground State}

As follows from the discussion of the properties of the eigenvalues of
the transfer matrix at the beginning of the previous section, and as shown
explicitly in Figure 1, all of the zeroes of the functions $\mathbb{T}
(\theta)$ and $\mathbb{T}(\theta+2\pi i)$ are in this case located in the
vicinity of the lines $\imag (\theta )=\pm 5\pi/6$ and $\imag (\theta )=
\pm 7\pi/6$ respectively. In particular, they do not contain any zeroes
in the strip $|\imag (\theta )|<\pi/3 $. Moreover, their ratio is real
and positive for all real $\theta$. Therefore the functions
\beq
\epsilon_0(\theta)=\log\mathbb{Y}_0(\theta), \qquad\epsilon_{1}(\theta
)=\log\mathbb{Y}_1(\theta)   \label{epsilons1}
\eeq
are real analytic for $|\imag (\theta )|<\pi/3 $. Taking into account
the asymptotics (\ref{eq:newasymp}), it follows that the functions
$\epsilon_0(\theta)-r\cosh\theta$ and $\epsilon_1(\theta)$ are regular
and bounded in the strip  $|\imag (\theta )|<\pi/3 $. Therefore 
one can  bring\cite{BazhanovLukyanovZamolodchikov4} the functional relation
(\ref{eq:newrel}) to the form
\beq
\epsilon_{j}(\theta )=\delta_{j,0}\, mR\,\cosh\theta +\sum_{k=0}^{1}
\Phi_{j,k}*L_{k}^{(-)}(\theta), \qquad j=0,1  \label{eq6:tbae1}
\eeq
where the kernel $\Phi_{jk}(\theta)$ and the function $L_{k}^{(\pm)}(\theta )$,
are respectively defined by
\beq
\Phi_{j,k}(\theta) =\left(\rule{0mm}{3.5mm}\delta_{j,k}-\delta_{j,k-1}-
\delta_{j,k+1}\right)\phi(\theta)    \label{eq6:bigphifn}
\eeq
\beq
L_{k}^{(\pm)}(\theta )=\log\left( 1+e^{\pm\epsilon_{k}(\theta )}\right) ,
\eeq
the function $\phi(\theta)$ is defined in (\ref{phi}), and the convolution
operator $*$ is defined by
\beq
f*g(\theta )=\int_{-\infty}^{\infty}f(\theta\minus\theta^{\prime})g(\theta^{
\prime}) d\theta^{\prime} . 
\eeq
Equations (\ref{eq6:tbae1}) are identical to the TBA integral equations
(\ref{eq:tbae1}). Similarly, the energy expression (\ref{eq:eofr2}) becomes
\beq
E_{0}(R)={m^2 R\,\log(m R_0)\over8\pi}-\frac{m}{2\pi}\int_{-\infty}^{\infty}
\cosh\theta\, L_{0}^{(-)}(\theta)\, d\theta  . \qquad  \label{eq6:energy1}
\eeq
which is also identical to the earlier TBA result (\ref{eq:eofr1}), at least 
up to some appropriate bulk term. Consequently, the conformal properties
associated with the scaling limit of these equations are the same as those
calculated in section 2.

\subsection{The TBA Equations for the First Excited State}

As shown in Figure 2, the location of the zeroes of $\mathbb{T}(\theta)$
and $\mathbb{T}(\theta\plus 2\pi i)$, is the same as above except that
both of these functions have two extra zeroes (\ref{eq:twozeroes}) at $\theta
=\pm i\pi$. These extra zeroes cancel out in (\ref{eq:newfunc}) resulting in
the functions $\mathbb{Y}_0(\theta)$ and $\mathbb{Y}_1(\theta)$ being real
and negative for all real $\theta$. Defining the functions
\beq
\epsilon_0(\theta)=\log\left(-\mathbb{Y}_0(\theta)\rule{0mm}{3.5mm}\right),
\qquad\epsilon_1(\theta)= \log\left(-\mathbb{Y}_1(\theta)\rule{0mm}{3.5mm}
\right),     \label{epsilons2}
\eeq
and proceeding as in the case of the ground state, one obtains from the
functional relation (\ref{eq:newrel}) and energy expression (\ref{eq:eofr2}),
the first excited state TBA equations
\begin{equation}
\epsilon_{j}(\theta )=\delta_{j,0}\, mR\,\cosh (\theta )+\sum_{k=0}^{1}
\Phi_{j,k}*\log\left(\rule{0mm}{4mm}1-e^{-\epsilon_k(\theta )}\,\right)
, \qquad j=0,1,   \label{eq:tbae2}
\end{equation}
where the function $\Phi_{j,k}(\theta)$ is defined in (\ref{eq6:bigphifn}),
and the corresponding first excited state energy
\begin{equation}
E_{1}(R)= {m^2 R\,\log(m R_0)\over8\pi}-\frac{m}{2\pi}\int_{-\infty
}^{\infty}\cosh\theta\,\log\left(\rule{0mm}{4mm}1-e^{-\epsilon_0(
\theta)}\,\right)\,\, d\theta .   \label{eq:eofr3}
\eeq
The leading short distance asymptotics of $E_1(R)$ determined by these
equations can be shown using the standard dilogarithm trick
\cite{TsvelickWiegmann} to be
\beq
E_1(R)\sim\frac{\pi}{10R}
\eeq
which is in agreement with the expected form $E_1(R)\sim-(\pi/6R)(c-24
\Delta_{1})$, where $\Delta_{1}=0$ is the second lowest conformal
dimension in the operator algebra associated with the minimal CFT model
$\mathcal{M}_{3,5}$. Equations (\ref{eq:tbae2}) and (\ref{eq:eofr3})
were previously conjectured in \cite{Ravanini3} using a different approach.

\subsection{The TBA Equations for the Second Excited State}

The location of the zeroes of $\mathbb{T}(\theta)$ and $\mathbb{T}(\theta\plus
2\pi i)$ in this case depends on the value of the parameter $r$. As described
in the previous section (and shown in Figures 3a and 3b), most of the zeroes
of these functions are the same as for the first excited state except that there
is an extra pair of zeroes. The positions of these extra zeroes depends upon
whether the parameter $r<r_{c}$ or $r>r_{c}$, where the critical value $r_{c}$
is specified in (\ref{eq6:critr}). For $r<r_{c}$, the extra zeroes of $\mathbb{T}
(\theta)$ lie on the real $\theta$ axis at the positions $\theta =\pm\alpha,
\; \alpha>0$. For $r>r_{c}$, the extra zeroes of $\mathbb{T}(\theta)$ lie on
the imaginary $\theta$ axis at the positions $\theta =\pm i\gamma ,\;\gamma
>0$. These two cases are described separately below.
\vspace{5mm}

\noindent\textbf{The case $\bmath{r\! <\! r_{c}}$:} As follows from the patterns
of zeros described above, the function $\mathbb{Y}_0(\theta)$ has two zeroes in
the strip $|\imag (\theta )|<\pi/3$ located on the real $\theta$ axis at $\theta=
\pm\alpha$, $\;\alpha>0$. Correspondingly, the function $\mathbb{Y}_1(\theta)$
has simple poles at these points. Moreover, the large $\theta$ asymptotics of
$\mathbb{Y}_0(\theta)$ and $\mathbb{Y}_1(\theta)$ are given by
(\ref{eq:newasymp}). This situation is very similar to that described in
Reference \cite{BazhanovLukyanovZamolodchikov4} for the first excited
state of the scaling Lee-Yang model. Following the discussion presented there,
it is useful to introduce the functions
\beq
\begin{array}{l}
\sigma_0(\theta,\alpha)=\tanh\left({3\over4}\,\left(\rule{0mm}{3.5mm}
\theta -\alpha\right)\right) \\ \displaystyle
\sigma_{1}(\theta,\eta)={{\cosh(\theta)\minus\cos(\eta)}\over
{\cosh(\theta)\plus\cos(\eta)}}\ {{\cosh(\theta)\minus\sin(\pi/6\minus
\eta)} \over {\cosh(\theta)\plus\sin(\pi/6\minus\eta)}} \rule{0mm}{8mm}
\end{array}
\eeq
which satisfy the equations
\beq\sigma_0 (\theta\plus i\pi/3)\,\sigma_0 (\theta\minus i\pi/3)=1,
\qquad \sigma_1 (\theta\plus i\pi/3,\eta)\,\sigma_1 (\theta\minus i
\pi/3,\eta)=\sigma_1 (\theta,\eta)\ .
\eeq
Now, defining the functions
\beq
\epsilon_0(\theta)=\log\mathbb{Y}_0(\theta)-\log\sigma(\theta,\alpha),
\qquad\epsilon_1(\theta)= \log\mathbb{Y}_1(\theta)+\log\sigma(\theta,\alpha),
\label{epsilons3}
\eeq
where
\beq
\sigma(\theta,\alpha) = \sigma_0(\theta,\alpha)\sigma_0(\theta,-\alpha),
\eeq
it then follows that the functions $\epsilon_0(\theta)-r\cosh\theta$ and
$\epsilon_1(\theta)$ are regular and bounded in the strip $|\imag\theta|
<\pi/3 $. Applying the above lemma one then obtains from
(\ref{eq:newrel}) the integral equations
\begin{equation}
\begin{array}{l}\displaystyle \epsilon_{0}(\theta )=r\,\cosh (\theta)+
\int_{-\infty}^\infty\!\phi(\theta\minus\theta')\left[\log\left(
\rule{0mm}{4mm}\sigma(\theta,\alpha)\plus e^{-\epsilon_0(\theta')}
\right)- \right.\\
\qquad\qquad\qquad\qquad\qquad\qquad\qquad\qquad\qquad\left. -\log\left(
\rule{0mm}{4mm}1\plus\sigma(\theta,\alpha) e^{-\epsilon_1(\theta')}
\right)\right] d\theta'\ \\
\epsilon_{1}(\theta )=r\cosh\theta-\epsilon_0(\theta)
\rule{0mm}{6mm} \end{array} \label{eq:tbae3}
\end{equation}
where $\phi(\theta)$ is defined in (\ref{phi}). The value of $\alpha$
which determines the position of the zeroes of $\mathbb{Y}_0(\theta)$,
is constrained  by the functional equation (\ref{Fnrel}). Indeed,
substituting $\theta=\alpha\pm i\pi/3$ in (\ref{Fnrel}) results in the
condition
\beq
\mathbb{Y}_0(\alpha\pm i\pi/3)=-1.
\eeq
Using the TBA equations (\ref{eq:tbae3}), and the explicit form
(\ref{phi}) of the kernel $\phi (\theta)$, one can rewrite this
condition as
\beqstar
r\sqrt{3}\sinh\alpha +{3\over\pi}\vpint{\cosh2(\theta\minus\alpha)\over
\sinh2(\theta\minus\alpha)}\left[\log\left(\rule{0mm}{4mm}\sigma(\theta,
\alpha)\plus e^{-\epsilon_0(\theta')}\,\right)-\right. \qquad\qquad
\qquad\qquad
\eeqstar
\beq
\qquad\qquad\qquad\left. -\log\left(\rule{0mm}{4mm}1\plus\sigma(\theta,\alpha
)e^{-\epsilon_1(\theta')}\,\right)\right] d\theta' =\pi\left(4I+2\arctan(
\sinh3\alpha)\rule{0mm}{4mm}\right)  \label{eq6:tbae3asub}
\eeq
where $I$ is some integer (arising when taking the logarithm of
(\ref{eq:tbae3})), and where ${\rm -}\kern -.92em\int_{-\infty}^\infty$
denotes the principal value of the singular integral. Numerical
calculations indicate that in this case $I=0$. Finally, the energy
expression (\ref{eq:eofr2}) in the case $r<r_{c}$ becomes
\bea
E_2(R) & \!\! =\!\! & {m^2 R\,\log(m R_0)\over8\pi}+{\sqrt{3}m}\cosh\alpha -
             \frac{m}{2\pi}\int_{-\infty}^{\infty}\!\cosh\theta\,\log\!\left|1+
             \mathbb{Y}_0(\theta)^{-1}\rule{0mm}{4mm}\right|\, d \theta
             . \nonumber \\
 & \!\! =\!\! & {m^2 R\,\log(m
R_0)\over8\pi}-\frac{m}{2\pi}\int_{-\infty}^{\infty} 
      \cosh\theta\,\log\left(\rule{0mm}{4mm}\sigma(\theta,\alpha)+ e^{-
       \varepsilon_0(\theta)}\,\right)\,\, d\theta, \qquad\qquad
       \label{eq6:energy3a}
\eea
Again, the leading short distance asymptotics of $E_{2}(R)$ can be
found through 
an appropriate modification of the dilogarithm trick \cite{KlumperPearce2,
KunibaNakanishi,BazhanovLukyanovZamolodchikov4}. It is not necessary to present
further details of this calculation here as they are well described in 
Appendix~C of Reference \cite{BazhanovLukyanovZamolodchikov4}. The final 
result is
\beq
E_{2}(R)\sim\frac{9\pi}{10R}
\eeq
which is in agreement with the expected form $E_{2}\sim -(\pi/6R)(c
\minus 24\Delta_{2})$, where $\Delta_{2}\! =\! 1/5$ is the third lowest
conformal dimension in the operator algebra associated with the
minimal CFT model ${\cal M}_{3/5}$.
\vspace{5mm}

\noindent\textbf{The case $\mathbf{r\! >\! r_{c}}$:} The consideration of
this case is very similar to the case $r<r_c$ described above. The function
$\mathbb{Y}_0(\theta)$ (and respectively $\mathbb{Y}_1(\theta)$) has two
complex conjugated zeroes (poles) at $\theta\! =\!\pm i\gamma$, $\;\gamma>0$.
Proceeding in the same manner as presented above, one obtains from
(\ref{eq:newrel})
\begin{equation}
\begin{array}{l} \displaystyle\epsilon_{0}(\theta )=r\,\cosh (\theta)+\log
\sigma_1(\theta,\gamma) +\phi *\left[ L_{0}^{(-)}(\theta )-L_{1}^{(-)}
(\theta )\right] \\
\epsilon_{1}(\theta )=r\cosh\theta-\epsilon_0(\theta) \rule{0mm}{6mm}
\end{array}
\label{eq6:tbae3b}
\end{equation}
where
\beq\label{epsilons4}
\epsilon_0(\theta)=\log\mathbb{Y}_0(\theta), \qquad\epsilon_1(\theta)= \log
\mathbb{Y}_1(\theta) .
\eeq
The value of $\gamma$ is here determined by the condition
\beq
\mathbb{Y}_0(\pm i(\gamma\minus\pi/3))=-1.  \label{eq6:tbae3bsub}
\eeq
The corresponding energy expression for $r>r_{c}$ is
\beq
E_2(R)={m^2 R\,\log(m R_0)\over8\pi}+{2m}\sin\left(\rule{0mm}{3.5mm}\pi/3
\minus\gamma\right) -\frac{m}{2\pi}\int_{-\infty}^{\infty}\!\cosh\theta\,\log
\left( 1+\mathbb{Y}_0(\theta)^{-1}\rule{0mm}{4mm}\right)\, d\theta. \qquad
\label{eq6:energy3b}
\eeq

\section{Numerical Results and Discussion}
\subsection{Numerical results}
In order to numerically determine the ground state and lowest two excited
state energy levels $E_{0}(R)$, $E_{1}(R)$ and $E_{2}(R)$ as a function of
the finite length $R$, it is first necessary to numerically solve the
respective systems of TBA integral equations for all required values of $R$. 
In the case of the ground state energy $E_{0}(R)$, the associated TBA 
equations (\ref{eq6:tbae1}) are easily solved at particular
values\footnote{We examined a number of different $R$ values in the range 
$10^{-4}<R<10$.} of $R$ using a simple iterative procedure. Using these 
solutions it is then straightforward to numerically integrate the energy 
expression (\ref{eq6:energy1}) to obtain $E_{0}(R)$ at these values of $R$.
\begin{figure}[t]
\centering
\vspace{5mm}
\includegraphics[scale=0.5,angle=-90]{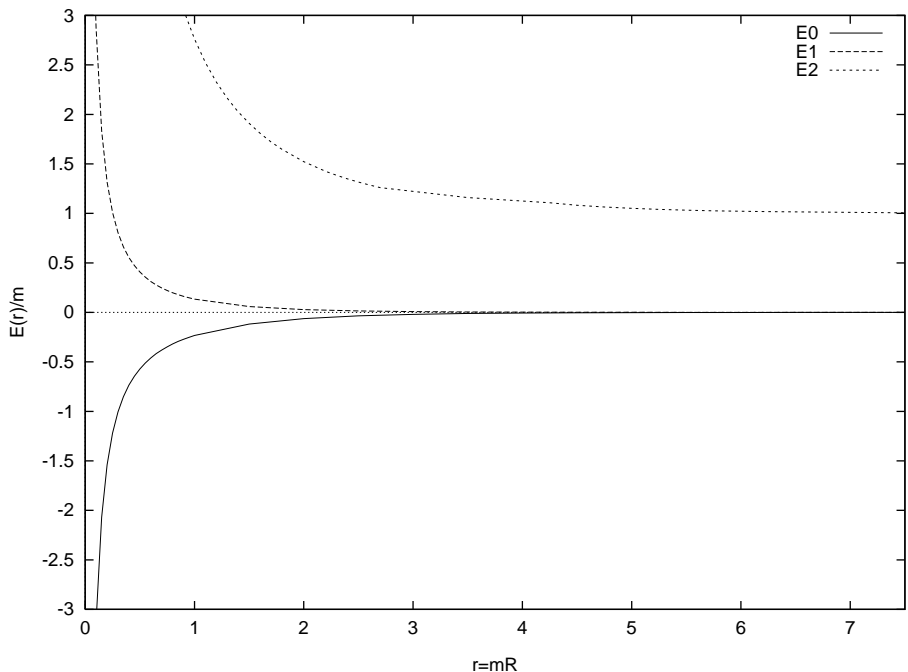}
\vspace{2.5mm}
\caption{The lowest three energy levels $E_0(R)$, $E_1(R)$ and $E_2(R)$ of the
$\mathcal{M}_{3,5}\pm\phi_{2,1}$ QFT obtained numerically from the energy
expressions (\ref{eq6:energy1}), (\ref{eq:eofr3}), (\ref{eq6:energy3a}) and
(\ref{eq6:energy3b}), after numerically solving the respective TBA integral 
equations (\ref{eq6:tbae1}), (\ref{eq:tbae2}), (\ref{eq:tbae3}) with the 
constraint (\ref{eq6:tbae3asub}), and (\ref{eq6:tbae3b}) with the constraint
(\ref{eq6:tbae3bsub}).}
\end{figure}
\begin{figure}[ht]
\centering
\vspace{5mm}
\includegraphics[scale=0.5,angle=-90]{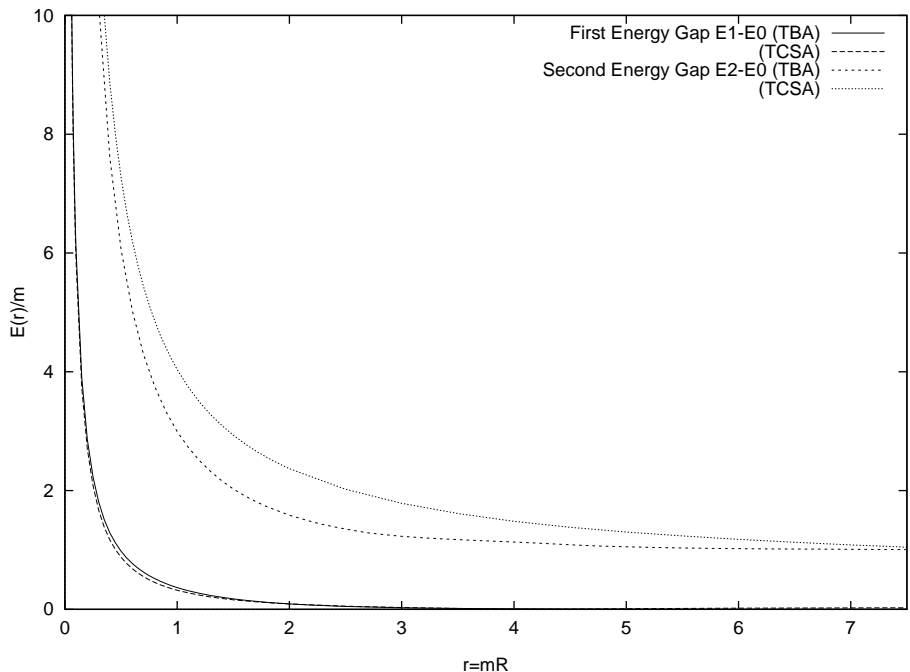}
\vspace{2.5mm}
\caption{The two lowest energy level differences $E_1(R)-E_0(R)$ and
$E_2(R)-E_1(R)$ obtained numerically from the TBA data shown in Figure 6.4
and presented respectively in Table 6.1 and 6.2, and from the TCSA method
\cite{Mussardo2} described in Section 5.4.1.}
\end{figure}

For the first excited state energy level $E_{1}(R)$ defined by 
(\ref{eq:eofr3}), the associated TBA equations (\ref{eq:tbae2}) can again 
be solved numerically at particular values of $R$ using a simple iterative 
procedure. The only significant difference in this case, is that there are 
convergence problems which can arise due to the fact that it is now 
possible for the argument of the logarithmic term in (\ref{eq:tbae2}) to 
become either very small, or even negative. Once a numerical solution of 
(\ref{eq:tbae2}) is known, it is again straightforward to numerically 
evaluate the integral in (\ref{eq:eofr3}) and hence obtain $E_{1}(R)$ for
the chosen values of $R$.
.

For the second excited state energy level $E_{2}(R)$, the situation is
much more complicated. In this case there are two distinct regions of 
$R$ defined respectively by the bounds $r<r_{c}$ and $r>r_{c}$, where 
the critical value $r_{c}$ is specified by (\ref{eq6:critr}). In fact, 
these two regions are described by quite different energy expressions 
and TBA equations.  For $r<r_{c}$, the energy level $E_{2}(R)$ is 
determined by the expression (\ref{eq6:energy3a}), which depends on the 
TBA equations (\ref{eq:tbae3}) and the associated constraint condition 
(\ref{eq6:tbae3asub}). Similarly, for $r>r_{c}$ the energy level 
$E_{2}(R)$ is determined by the expression (\ref{eq6:energy3b}), which 
depends on the TBA equations (\ref{eq6:tbae3b}) and the associated 
constraint condition (\ref{eq6:tbae3bsub}). In both of these cases, 
accounting for the additional constraint equations as well as the 
convergence problems (which are similar in nature to those described 
above) present some extra difficulties for the numerical solution of
the two sets of TBA equations.  However, a more significant problem 
in this case is that all of the aforementioned numerical strategies 
breakdown in the region around the critical value $r_{c}$ of the 
scaling parameter $r$. This problem is avoided here through the use 
of an interpolating function to describe the energy level
$E_{2}(R)$ in the vicinity of the crossover between these two 
descriptions.
{\def\baselinestretch{1.0}
\begin{table}[t]
\centering
\begin{tabular}{|l||l|l|}
\hline
$\hfill mR\hfill$ & \hfil $E_{1}\minus E_{0}$ (TBA)\hfill
 & \hfil (TCSA)\hfill \\ \hline \hline
$0.00001$ & $62831.56445476$ & $62831.35411$ \\
$0.000025$ & $25132.45254369$ & $25132.24227$ \\
$0.00005$ & $12566.08193960$ & $12565.87166$ \\
$0.000075$ & $8377.291734939$ & $8377.081465$ \\
$0.0001$ & $6282.896637100$ & $6282.686370$ \\
$0.00025$ & $2512.985451883$ & $2512.775232$ \\
$0.0005$ & $1256.348396107$ & $1256.138248$ \\
$0.00075$ & $837.4693800061$ & $837.2593044$ \\
$0.001$ & $628.0298740896$ & $627.8198712$ \\
$0.0025$ & $251.0387832873$ & $250.8292145$ \\
$0.005$ & $125.3751232181$ & $125.1662761$ \\
$0.0075$ & $83.48726742289$ & $83.27913960$ \\
$0.01$ & $62.54336272361$ & $62.33595193$ \\
$0.025$ & $24.84453166641$ & $24.64137450$ \\
$0.05$ & $12.27863838885$ & $12.08239070$ \\
$0.075$ & $8.090337319392$ & $7.900781170$ \\
$0.1$ & $5.996443118481$ & $5.813368363$ \\
$0.25$ & $2.229783044794$ & $2.081577717$ \\
$0.5$ & $0.9794797481711$ & $0.876706700$ \\
$0.75$ & $0.5680114390827$ & $0.498611616$ \\
$1.0$ & $0.3669407772837$ & $0.321935555$ \\
$2.5$ & $0.05008836227296$ & $0.0531794399$ \\
$5.0$ & $0.00286684507277$ & $0.0101268211$ \\
$7.5$ & $0.00018876702243$ & $0.0276849431$ \\
\hline
\end{tabular}
\caption{A comparison of some numerical results for the lowest
energy gap $E_{1}(R)\minus E_{0}(R)$ evaluated from the TBA approach
using equations (\ref{eq6:tbae1}), (\ref{eq6:energy1}), (\ref{eq:tbae2})
and (\ref{eq:eofr3}), and from the TCSA method \cite{Mussardo2} described
in Section 5.4.1.}
\end{table}}

{\def\baselinestretch{1.0}
\begin{table}[ht]
\centering
\begin{tabular}{|l||l|l|}
\hline
$\hfill mR\hfill$ & \hfil $E_{2}\minus E_{0}$ (TBA)\hfill
 & \hfil (TCSA)\hfill \\ \hline \hline
$0.001$ & $3141.370972547$ & $3142.779881$ \\
$0.002$ & $1570.806398677$ & $1571.982982$ \\
$0.004$ & $785.0624826259$ & $786.5836788$ \\
$0.006$ & $523.2879477632$ & $524.7831553$ \\
$0.008$ & $392.3470761712$ & $393.8823300$ \\
$0.01$ & $313.9321721220$ & $315.3413864$ \\
$0.02$ & $156.7885662688$ & $158.2561799$ \\
$0.04$ & $78.24246832636$ & $79.70552037$ \\
$0.06$ & $52.06938686002$ & $53.51512707$ \\
$0.08$ & $38.99405029948$ & $40.41507268$ \\
$0.1$ & $31.13990032524$ & $32.55135753$ \\
$0.2$ & $15.44209616262$ & $16.79944053$ \\
$0.4$ & $7.618391854072$ & $8.876100399$ \\
$0.6$ & $5.028671954372$ & $6.206070558$ \\
$0.8$ & $3.750083921383$ & $4.856641130$ \\
$1.0$ & $2.994096980864$ & $4.038850086$ \\
$2.0$ & $1.585738482126$ & $2.367628184$ \\
$2.5$ & $1.352089391730$ & $2.021744859$ \\
$4.0$ & $1.151471437505$ & $1.482792705$ \\
$5.0$ & $1.053476286910$ & $1.302407192$ \\
$6.0$ & $1.022056716644$ & $1.178501191$ \\
$7.5$ & $1.006154507587$ & $1.046907854$ \\
\hline
\end{tabular}
\caption{A comparison of some numerical results for the second lowest
energy gap $E_{2}(R)\minus E_{0}(R)$ evaluated from the TBA approach
using equations (\ref{eq6:tbae1}), (\ref{eq6:energy1}), (\ref{eq:tbae3}),
(\ref{eq6:tbae3asub}), (\ref{eq6:energy3a}), (\ref{eq6:tbae3b}),
(\ref{eq6:tbae3bsub}) and (\ref{eq6:energy3b}), and from the TCSA method
\cite{Mussardo2} described in Section 5.4.1.}
\end{table}}

In Figure 4, the energy levels $E_{0}(R)$, $E_{1}(R)$ and $E_{2}(R)$
derived from the TBA methods outlined above, are plotted as functions of
the scaling length $r=mR$.  A comparison of these excited state TBA 
results can also be made with the known TCSA results \cite{Mussardo2}. 
In particular, it is most useful to compare the energy level differences 
$E_1(R)\! -\! E_0(R)$ and $E_2(R)\! -\! E_0(R)$ obtained from both of 
these approaches. These energy level differences are plotted in Figure 5. 
Some particular values of these energy level differences are also
presented in Tables 1 and 2. It is quite obvious from Figure 5 that the agreement
between the TBA and TCSA results is extremely good for the first energy level
difference, but is somewhat poorer for the second energy level difference
(especially at intermediate values of $r$). A total  agreement with TCSA is
not really anticipated since the dimension 
of the perturbation in (\ref{action}) $\Delta=3/4$ exceeds the value 
$\Delta=1/2$ above which the TCSA method requires modifications due to
divergence problems.
We refer the reader to \cite{lassig} where 
these problems of the TCSA method are more thoroughly discussed.

\subsection{Massless case}
Consider now the asymptotic expansions of the operator ${\mathbb T}(\theta)$
at large values of $\theta$. For simplicity restrict ourselves to the
CFT case by replacing the $r\,\cosh(\theta)$ term in (\ref{Fnrel}),
(\ref{eq:newfunc}) by 
$\frac{1}{2}\,r\,e^\theta$. 
The equation (\ref{Fnrel}) can now be written as
 \begin{equation}
e^{-{r\over8}e^\theta}{{\mathbb T}(\theta+i\pi/3){\mathbb T}
(\theta-i\pi/3)\over
{\mathbb T}(\theta)}=1+{\mathbb Y}_0^{-1}(\theta)\label{conftt}
\end{equation}

The ground state eigenvalue ${\mathbb T}^{(0)}(\theta)$ of ${\mathbb
T}(\theta)$  
does not have zeroes in the strip $|\imag \theta|<\pi/3$. Therefore, with an  
account of the asymptotics (\ref{eq:bigtasymp}), the equation
(\ref{conftt}) implies 
\begin{equation}
\log{\mathbb T}^{(0)}(\theta)=
-{\sqrt{3} r\over 8 \pi}\theta e^\theta+ {\mathbb C} e^\theta
+\phi\ast \log (1+e^{-\epsilon_0(\theta)})
\label{confexp}\end{equation}
where ${\mathbb C}$ is an arbitrary constant and 
$\epsilon_0(\theta)$ determined by the ``massless'' version of
(\ref{eq6:tbae1}) with the  $r\,\cosh(\theta)$ term replaced by
$\frac{1}{2}\,r\,e^\theta$. 
 
Expanding the kernel (\ref{phi}) in a series 
\begin{equation}
\phi(\theta) = -{2\over \pi}\
\sum_{n=1}^{\infty}\sin{{2\pi(n+1)}\over
3}\  e^{(1-2n)\theta}\ .
\end{equation}
one gets from (\ref{confexp}) 
\begin{equation}	
\log{\mathbb T}^{(0)}(\theta)
=-{\sqrt{3} r\over 8 \pi}\theta e^\theta+ {\mathbb C} e^\theta-{2\over \pi}\
\sum_{n=1}^{\infty}\sin{{2\pi(n+1)}\over
3}\ \chi_n  e^{(1-2n)\theta}\ .\label{confexpp}
\end{equation}	
where
\begin{equation}	
\chi_n=\int_{-\infty}^{\infty}e^{(2n-1)\theta)} 
(1+e^{-\epsilon_0(\theta)})d\theta\label{chidef}
\end{equation}	
The numerical values of these coefficients can be compared with 
the the exact results of \cite{BHK} where CFT's with extended ${\cal W}_3$
symmetry were studied.
To facilitate this comparison let us introduce a new variable 
\begin{equation}
t^2=\frac{3r e^\theta}{8\pi a^2}
\end{equation}
where $a$ is a normalisation constant and rewrite the first terms of the 
expansion (\ref{confexpp}) as
\begin{equation}
{\mathbb T}^{(0)}(t)=-{2\over\sqrt{3}}(at)^2\log t+{\mathbb C}\,t^2-
{3\sqrt{3}\over 2}\,(at)^{-2}\, I^{(0)}_1+
{\frac {32805\, \sqrt {3}}{43472}} (at)^{-10}\, I^{(0)}_5+\ldots\label{tas56}
\end{equation}	
According to \cite{BHK} the quantities  $I^{(0)}_1$ and $I^{(0)}_5$ are the
vacuum eigenvalues of the local integrals of motion in a particular 
highest weight module of the ${\cal W}_3$ algebra with the value of
the central charge $c=6/5$ (see \cite{BHK} for further details). Their 
exact values \cite{BHK}
\begin{equation}
I^{(0)}_1=-\,\frac{1}{20}, \qquad I^{(0)}_5=-\,\frac{121}{10500}
\end{equation}
are in good agreement with numerical values obtained from
(\ref{chidef})
\begin{equation} 
I^{(0)}_1=-0.499999999599\ 10^{-1}, \qquad I^{(0)}_5=-0.115238095240\ 10^{-1}
\end{equation}

Similar calculations for the first excited state of Sect. 4.2 in the
``massless'' limit give 
\begin{equation} 
I^{(1)}_1=0.500000000002\ 10^{-1}, \qquad I^{(1)}_5=0.497619047634\  10^{-2}
\end{equation}
which again in a good agreement with the corresponding exact values of
\cite{BHK} 
\begin{equation} 
I^{(1)}_1=\frac{1}{20}, \qquad I^{(1)}_5=\frac{209}{42000}
\end{equation}

\subsection{Functional Equations}

Finally note, that the functional 
equation (\ref{eq:fusionreln}) can be obtained from 
the general set of of $A^{(2)}_2$ fusion relations \cite{KunibaSuzuki} 
\beq
\bmath{T}_{n}\bigg(u+\lambda-\frac{\pi}{2}\bigg)\ 
\bmath{T}_{n}\bigg(u-\lambda+\frac{\pi}{2}\bigg)=
\bmath{T}_{n-1}(u)\ \bmath{T}_{n+1}(u)+f_{n+1}\bigg(u-\frac{3\lambda}{2}\bigg)
\ \bmath{T}_{n}
\bigg(u+\frac{\pi}{2}\bigg)\label{fusion}
\eeq
where
\beq
\bmath{T}_{-1}(u)\equiv0,\qquad \bmath{T}_0=(-1)^N\, 
f_0\bigg(u-\frac{3\lambda}{2}\bigg),\qquad \bmath{T}_1(u)\equiv\bmath{T}(u),
\eeq
\beq
f_{n}(u)=\left(p^{-\frac{1}{2}}
 \theta_1(u-v_n)\theta_1(u+v_n)\right)^{N},\qquad
v_n=\frac{\pi}{4}((2n+1)g+1)
\eeq
where the theta function $\theta_1(u)=\theta_1(u,p)$ defined in
(\ref{eq:thetafns}) 
and
\beq
g=1-\frac{2\lambda}{\pi}
\eeq
We would like to stress that the relation 
(\ref{fusion}) holds for arbitrary values 
of the parameters $\lambda$ and $\omega$ in (\ref{eq:eigval}),
where it can be considered just as a definition the ``fused'' 
transfer matrices $\bmath{T}_n(u)$, $n\ge2$, in terms of
$\bmath{T}_1(u)\equiv\bmath{T}(u)$.
The eigenvalue expression
(\ref{eq:eigval}) and the Bethe Ansatz equations (\ref{BAE})
ensure that all the higher 
transfer matrices $\bmath{T}_n(u)$, $n\ge2$, are entire functions of
the variable $u$ as well as $\bmath{T}(u)$. 
For some special  values of of the parameters $\lambda$ and $\omega$
the infinite set of relation (\ref{fusion}) 
truncates and becomes a system of functional equations for 
a finite number of the transfer matrices $\bmath{T}_n(u)$.
Let us a introduce a new variable  
\beq
t=p^{-1} e^{2i(u-3\lambda/2)}
\eeq
and rescale the transfer matrices as
\beq
\bmath{T}_n(u)=(p t)^{-N}{\bf T}_n(t).
\eeq
The relation (\ref{fusion}) then becomes
\beq
{\bf T}_{n}(t\, q)\ 
{\bf T}_{n}(t\, q^{-1})=
{\bf T}_{n-1}(t)\ {\bf T}_{n+1}(t)+\phi_{n+1}(t)
\ {\bf T}_{n}(-t)\label{fusion1}
\eeq
with
\beq
q=e^{i\pi g}
\eeq
and 
\beq
\phi_n(t)=\big[\phi(i t q^{n+\frac{1}{2}})\phi(-i t
q^{-n-\frac{1}{2}})\big]^N,\qquad
\phi(t)=\sum_{k=-\infty}^{\infty}(-1)^n p^{n^2} t^n
\eeq
Now it is easy to see that if the the scalar factor
$\phi_{n+1}(t)$ is omitted 
the relation (\ref{fusion1}) coincide with a particular 
case (corresponding to the $A^{(2)}_2$ reduction) of the more 
general $A^{(1)}_2$ fusion relations considered in ref.\cite{BHK} (see
eqs.(5.14), 
(6.47), (6.48) therein). Note, that all reasonings of \cite{BHK} can me
modified to take into account the above scalar factor $\phi_{n+1}(t)$
in (\ref{fusion1}).  
Then using the determinant expressions (5.1),
(5.2) of ref.\cite{BHK} (see also the discussion in Sect.8.1 of that paper) 
one can show that if
\beq
q^{2(m+2)}=1,\qquad{\rm and}\qquad \omega^{m+2}=1,
\eeq 
for some integer $m$ then
\beq
{\bf T}_m(t)\equiv0,\qquad {\bf T}_{m-k-1}(t)={\bf
T}_{k}(-q^{m+2}t),\qquad k=0,1,\ldots,m-1.\label{reduc}
\eeq
The equation (\ref{eq:fusionreln}) corresponds to a
particular case 
of this reduction 
\beq
m=4,\qquad \lambda=\pi/12,\qquad q=e^{5\pi i/6}
\eeq
Another simple case is 
\beq
m=4, \qquad \lambda=\pi/6,\qquad q=e^{2\pi/3},
\eeq
where (\ref{fusion1}) and (\ref{reduc}) imply (assuming $N$ even)
\beq
{\bf T}(tq^{\frac{1}{2}})\ {\bf
T}(tq^{-\frac{1}{2}})=\big(\phi_0(-t)+\phi_2(-t)\big)
\ {\bf T}(t)\label{fusion3}
\eeq
The leading $p\to0$ singularity of the free energy of the dilute $A_2$ lattice
model of Sect.3 with this value of $\lambda=\pi/6$ is \ $\log k(u) \sim
p^2\log p \sin(2u)$,\  which determines to the value of the correlation
length exponent $\nu=1$. Therefore the scaling limit of the model 
corresponds to
\beq
N\to\infty,\qquad p\to0, \qquad r=4\sqrt{3} pN ={\rm finite}
\eeq
where  the equation (\ref{fusion3}) becomes
\beq
{\bf T}\bigg(\t+\frac{i\pi}{3}\bigg)\ {\bf
T}\bigg(\t-\frac{i\pi}{3}\bigg)=2\cosh\bigg(\frac{r}{2}\cosh(\t)\bigg)\
{\bf T}(\t)
\eeq
with $\t=\log t$. The massless version of this equation is related 
to a certain $c=-2$ CFT with extended ${\cal  W}_3$ symmetry
\cite{BHK} (Sect.6.3.2 therein).

%%%%%%%%%%%%%%%%%%%%
% Acknowledgements %
%%%%%%%%%%%%%%%%%%%%

\section*{Acknowledgements}

The authors are grateful to S. O. Warnaar and P. Pearce for interesting 
discussions.

\renewcommand{\thesection}{\Alph{section}}

\section*{Appendix A}
\setcounter{section}{1}
\setcounter{equation}{0}

The Boltzmann weights of the off-critical dilute-$A_{M}$ models are given by 
\cite{WarnaarNienhuisSeaton,WarnaarPearceSeatonNienhuis}
\bea
\boltzmannweight{a}{a}{a}{a}{u}{}{}\! & = & \rho\frac{\te(6\lambda\minus u)
      \te(3\lambda\plus u)}{\te(6\lambda)\te(3\lambda)} \nonumber \\
 & - & \!\!\rho\left(\frac{S(a\plus 1)}{S(a)}\frac{\tv(2a\lambda\minus 5\lambda
      )}{\tv(2a\lambda\plus\lambda)}+\frac{S(a\minus 1)}{S(a)}\frac{\tv(2a
      \lambda\plus 5\lambda )}{\tv(2a\lambda\minus\lambda)}\right)\frac{\te(u)
      \te(3\lambda\minus u)}{\te(6\lambda)\te(3\lambda)}   \nonumber \\
\boltzmannweight{a\pm 1}{a}{a}{a}{u}{}{}\! & = &
\boltzmannweight{a}{a}{a}{a\pm  
      1}{u}{}{}= \rho\frac{\te(3\lambda\minus u)\tv(\pm
2a\lambda\plus\lambda\minus  
      u)}{\te(3\lambda)\tv(\pm 2a\lambda\plus\lambda)} \nonumber \\
\boltzmannweight{a}{a}{a\pm 1}{a}{u}{}{}\! & = & \boltzmannweight{a}{a\pm 1}{a}
      {a}{u}{}{}= \rho\left(\frac{S(a\!\pm\!1)}{S(a)}\right)^{1/2}
\frac{\te(u)\tv 
      (\pm 2a\lambda\minus 2\lambda\plus u)}{\te(3\lambda)\tv(\pm
2a\lambda\plus 
      \lambda)} \nonumber \\
\boltzmannweight{a}{a\pm 1}{a}{a\pm 1}{u}{}{}\! & = & \boltzmannweight{a\pm 1}
      {a\pm 1}{a}{a}{u}{}{} =\nonumber \\
 & = & \!\!\rho\left(\frac{\tv(\pm 2a\lambda\plus 3\lambda)\tv(\pm
2a\lambda\minus 
      \lambda)}{\tv^2(\pm 2a\lambda\plus\lambda)} \right)^{1/2} \frac{\te(u) 
      \te(3\lambda\minus u)}{\te(2\lambda)\te(3\lambda)} \nonumber \\
\boltzmannweight{a\pm 1}{a}{a}{a\mp 1}{u}{}{}\! & = &
\rho\frac{\te(2\lambda\minus  
      u)\te(3\lambda\minus u)}{\te(2\lambda)\te(3\lambda)} \nonumber \\
\boltzmannweight{a}{a\mp 1}{a\pm 1}{a}{u}{}{}\! & = &
\!\!-\rho\left(\frac{S(a\minus 
      1)S(a\plus 1)}{S^2(a)}\right)^{1/2} \frac{\te(u)\te(\lambda\minus u)}{\te
      (2\lambda)\te(3\lambda)}   \nonumber \\
\boltzmannweight{a}{a\pm 1}{a\pm 1}{a}{u}{}{}\! & = &
\!\!\rho\frac{\te(3\lambda 
      \minus u)\te(\pm 4a\lambda\plus 2\lambda\plus u)}{\te(3\lambda)\te(\pm 
      4a\lambda\plus 2\lambda)}+\rho\frac{S(a\!\pm\! 1)}{S(a)}
\frac{\te(u) \te(\pm  
      4a\lambda\minus \lambda\plus u)}{\te(3\lambda)\te(\pm 4a\lambda\plus 2
      \lambda)} \nonumber \\
 & = & \!\!\rho\frac{\te(3\lambda\plus u)\te(\pm 4a\lambda\minus
4\lambda\plus u)}  
      {\te(3\lambda)\te(\pm 4a\lambda\minus 4\lambda)} \nonumber \\
 & + & \!\!\rho\left(\frac{S(a\!\mp\!
1)}{S(a)}\frac{\te(4\lambda)}{\te(2\lambda)} 
      -\frac{\tv(\pm 2a\lambda\minus 5\lambda)}{\tv(\pm
2a\lambda\plus\lambda)}  
      \right)\frac{\te(u)\te(\pm 4a\lambda\minus \lambda\plus
u)}{\te(3\lambda)  
      \te(\pm 4a\lambda\minus 4\lambda)} \nonumber \\
S(a) \!\!& = & \!\!(-)^{\displaystyle
a}\;\frac{\te(4a\lambda)}{\tv(2a\lambda)} \,  
\nonumber \\
\rho \!\!& = & \!\! h(2\lambda) h(3\lambda) . \nonumber
\eea

The standard elliptic functions of variable $u$ and nome $p=e^{-\tau}$ are
defined as
\beq
\te (u)\equiv \te (u;p)=2\,\sum_{n=0}^{\infty}(-1)^n
p^{(2n+1)^2/4}\sin [(2n+1)u] 
\label{eq:thetafns}
\eeq
\beq
\tv (u)\equiv \tv (u;p)=1+2\,\sum_{n=1}^{\infty}(-1)^n p^{n^2}\cos (2nu) .
\eeq
These have the useful (quasi-) periodicity properties
\beq
\vartheta_{1}(u\plus\pi ;p)=-\vartheta_{1}(u;p), \qquad\qquad
\vartheta_{1}(u\plus  
i\tau ;p)=-p^{-1}e^{-2iu}\vartheta_{1}(u;p)
\eeq
and
\beq
\vartheta_{4}(u\plus\pi,p)=\vartheta_{4}(u,p), \qquad\qquad
\vartheta_{4}(u\plus 
i\tau, p)=-p^{-1}e^{-2iu}\vartheta_{4}(u,p) .
\eeq

%%%%%%%%%%%%%%%%
% Bibliography %
%%%%%%%%%%%%%%%%

%{\def\baselinestretch{0.8}

%}

\end{document}